\documentclass[aps,prd,amsmath,amssymb,superscriptaddress,preprintnumbers,twocolumn,nofootinbib,10pt]{revtex4-1}
\pdfoutput=1
\usepackage{graphicx}
\usepackage{subfigure}
\usepackage{dcolumn}
\usepackage{bm}
\usepackage{amssymb}
\usepackage{latexsym}
\usepackage{booktabs}
\usepackage{amsmath}
\usepackage{multirow}
\usepackage{url}
\usepackage{footnote}
\usepackage{float}
\usepackage{threeparttable}
\usepackage[colorlinks=true, linkcolor=blue, citecolor=blue]{hyperref}
\usepackage[bottom]{footmisc}

\usepackage[normalem]{ulem}
\usepackage{color}
\usepackage{array}
\usepackage{enumerate}
\usepackage{adjustbox}

\usepackage{makecell}
\usepackage{diagbox}
\usepackage{epstopdf}
\usepackage{epsfig}
\usepackage{longtable}
\usepackage{supertabular}
\usepackage{algorithm}
\usepackage{pifont}
\usepackage{algorithmic}
\usepackage{changepage}
\usepackage{setspace}

\begin{document}

\title{Constraining neutrino masses with localized fast radio bursts: Current bounds and future prospects}

\author{Yin-Long Cao}\email{caoyinlong@stumail.neu.edu.cn}
\affiliation{Liaoning Key Laboratory of Cosmology and Astrophysics, College of Sciences, Northeastern University, Shenyang 110819, China}

\author{Ji-Guo Zhang}\email{zhangjiguo@stumail.neu.edu.cn}
\affiliation{Liaoning Key Laboratory of Cosmology and Astrophysics, College of Sciences, Northeastern University, Shenyang 110819, China}

\author{Guo-Hong Du}\email{duguohong@stumail.neu.edu.cn}
\affiliation{Liaoning Key Laboratory of Cosmology and Astrophysics, College of Sciences, Northeastern University, Shenyang 110819, China}

\author{Tian-Nuo Li}\email{litiannuo@ucas.ac.cn}
\affiliation{School of Fundamental Physics and Mathematical Sciences, Hangzhou Institute for Advanced Study, UCAS, Hangzhou 310024, China}

\author{Jing-Fei Zhang}\email{jfzhang@mail.neu.edu.cn}
\affiliation{Liaoning Key Laboratory of Cosmology and Astrophysics, College of Sciences, Northeastern University, Shenyang 110819, China}
 
\author{Xin Zhang}\thanks{Corresponding author}\email{zhangxin@mail.neu.edu.cn}
\affiliation{Liaoning Key Laboratory of Cosmology and Astrophysics, College of Sciences, Northeastern University, Shenyang 110819, China}
\affiliation{MOE Key Laboratory of Data Analytics and Optimization for Smart Industry, Northeastern University, Shenyang 110819, China}
\affiliation{National Frontiers Science Center for Industrial Intelligence and Systems Optimization, Northeastern University, Shenyang 110819, China}

\begin{abstract}
The dispersion-measure (DM)--redshift relation of fast radio bursts (FRBs) provides independent low-redshift cosmological information for neutrino-mass inference. In this work, we perform joint analyses with cosmic microwave background (CMB) data, using 120 localized FRBs with reliable host-galaxy redshifts and larger mock samples, to investigate the effects of sample size and the dark-energy model on constraints on the sum of neutrino masses, $\sum m_\nu$. In the $\Lambda$CDM+$\sum m_\nu$ model, CMB+FRB gives $\sum m_\nu<0.21\,\mathrm{eV}$, substantially relaxed relative to the current CMB+DESI result. With 1000 simulated events, CMB+FRB1(mock) gives an upper limit of $0.12\,\mathrm{eV}$, which decreases to $0.035\,\mathrm{eV}$ for CMB+FRB2(mock) with 10000 events. The latter is more stringent than the CMB+DESI limit, demonstrating the potential of large localized FRB samples to constrain neutrino mass. When the model is extended to $w_0w_a$CDM+$\sum m_\nu$, both the observed and mock samples yield upper limits of approximately $0.3$--$0.4\,\mathrm{eV}$, indicating that the assumed dark-energy model remains a dominant factor in these constraints. As localized FRB samples grow and the modeling of the individual DM contributions becomes more accurate, FRBs may provide further independent low-redshift information for constraining neutrino mass.
\end{abstract}

\maketitle

\section{INTRODUCTION}\label{sec1}

Neutrino oscillation experiments have established that neutrinos have nonzero masses, providing direct evidence for physics beyond the Standard Model of particle physics~\cite{Super-Kamiokande:1998kpq,SNO:2002tuh}. Solar and atmospheric neutrino oscillation experiments have measured the mass-squared differences between mass eigenstates, but have not yet fully determined the hierarchies of the three masses \cite{ParticleDataGroup:2024cfk}. Two possibilities remain: normal hierarchy (NH), with $m_1<m_2<m_3$, and inverted hierarchy (IH), with $m_3<m_1<m_2$. Here, $m_i$ denotes the mass of the $i$th neutrino mass eigenstate. The minimum sums of neutrino masses corresponding to these two hierarchies are approximately $0.06\,\mathrm{eV}$ and $0.10\,\mathrm{eV}$, respectively \cite{Esteban:2020cvm,deSalas:2020pgw}, but oscillation experiments alone cannot determine the absolute neutrino-mass scale. Tritium $\beta$-decay experiments provide a direct way to probe this scale. The KATRIN experiment constrains the effective electron-antineutrino mass to $m_\beta<0.45\,\mathrm{eV}$ at 90\% confidence \cite{KATRIN:2024cdt}.

Compared with current direct laboratory measurements, cosmological observations can place stronger constraints on the absolute neutrino-mass scale under standard cosmological-model assumptions. This constraining power arises from the effects of cosmic background neutrinos on the expansion history and structure growth. Under the standard thermal history, the neutrino abundance has a well-defined theoretical prediction, so their masses are directly related to their contribution to the cosmic energy density. Neutrinos behave predominantly as relativistic radiation in the early universe and contribute to the matter density after becoming nonrelativistic; both the timing of this transition and their energy density in the late universe depend on their masses. Meanwhile, neutrino free streaming suppresses structure growth below the corresponding free-streaming scale~\cite{Lesgourgues:2006nd}. High-precision cosmic microwave background (CMB) and large-scale-structure observations in the late universe can probe these effects, thereby placing stringent constraints on the sum of neutrino masses, $\sum m_\nu$  \cite{Zhang:2014nta,TopicalConvenersKNAbazajianJECarlstromATLee:2013bxd,Zhang:2014dxk}. In the spatially flat $\Lambda$ cold dark matter ($\Lambda$CDM) model, Planck 2018 CMB data combined with baryon acoustic oscillation (BAO) measurements give a 95\% upper limit of $\sum m_\nu < 0.12\,\mathrm{eV}$ \cite{Planck:2018vyg}; the Dark Energy Spectroscopic Instrument (DESI) Data Release 2 (DR2) BAO measurements combined with the CMB tighten this limit to approximately $0.064\,\mathrm{eV}$, approaching the minimum mass sum allowed by NH \cite{DESI:2025zgx}. In addition, existing studies have constrained and forecast neutrino properties using weak-lensing statistics, cross-correlations between galaxy clustering and CMB lensing, and 21 cm neutral-hydrogen intensity mapping \cite{Coulton:2018ebd,Zhang:2019ipd,Tanseri:2022zfe}.

Cosmological constraints on neutrino masses are highly sensitive to the assumed cosmological model. Our previous studies have shown that the choice of dark-energy model is one of the dominant factors affecting these constraints~\cite{Zhang:2015uhk,Zhao:2016ecj,Zhang:2017rbg}. Owing to the strong parameter degeneracies between the dark-energy equation of state and neutrino mass, different dark-energy evolution histories can substantially alter the upper bound on the sum of neutrino masses. Current DESI BAO measurements, combined with CMB and type Ia supernova (SN) data, suggest dynamical dark energy, with a statistical preference over $\Lambda$CDM reaching $4.2\sigma$~\cite{DESI:2025zgx}. These findings have prompted extensive discussion of the nature of dark energy~\cite{Li:2024qso,Giare:2024gpk,Jiang:2024xnu,Yang:2024kdo,Giare:2024smz,Ye:2024ywg,Liu:2024yib,Wolf:2025jed,Pang:2025lvh,Giare:2024oil,Liu:2025mub,Cheng:2025lod,Ozulker:2025ehg,Li:2025ops,Khoury:2025txd,Li:2025ula,Smith:2025icl,Li:2025dwz,Lemos:2025bgy,You:2025uon,Ling:2025lmw,Liu:2025myr,Li:2025htp,Li:2025muv,Wu:2025wyk,Wu:2024faw,Wang:2024dka,Li:2025vuh,Specogna:2025guo,Fazzari:2025lzd,Zhang:2025lam,Song:2025bio,Lee:2025pzo,Santos:2025wiv,Kessler:2025kju,Li:2025owk,Li:2026hwq,Li:2026xaz,Turner:2026vzp,Wang:2026wrk,Wang:2025znm,Yang:2025gaz,Wang:2026vqw,Yang:2026qmz,Pan:2025qwy,Yang:2026yaq,Du:2025xes,Pedrotti:2026ncd,Yao:2025twv,Yang:2026vti,Zhou:2026dti,Xu:2026sbw,Zhang:2026axl,Jiang:2026ngo,Ling:2026awm,Zhou:2026ive,Li:2026mbh,Li:2026jgg,Sun:2026kig,Du:2026qtq,Du:2026cly,Du:2025csv,Yang:2026qon,Pulido-Hernandez:2026hcs,Wang:2026gvg,Kibris:2026cqq,Zheng:2026tpt,Feng:2026pzs}; for recent reviews, see, e.g., Refs.~\cite{CosmoVerseNetwork:2025alb,Giare:2025pzu,Zhang:2026zbs,Cai:2025mas,Anchordoqui:2026hys,Li:2026asg,Giare:2026oti}. Specifically, within the $w_0w_a$CDM framework~\cite{Chevallier:2000qy,Linder:2002et}, DESI-based analyses favor a dark-energy equation of state that evolves from $w<-1$ in the past to $w>-1$ at present. As indicated by our previous studies, such an evolution can relax the upper bound on the neutrino mass sum~\cite{Zhang:2015uhk,Zhao:2016ecj,Zhang:2017rbg}. The joint analysis of DESI DR2 BAO and CMB data yields $\sum m_\nu<0.16\,\mathrm{eV}$~\cite{DESI:2025zgx}.

However, DESI BAO and CMB measurements show a discrepancy of $2$--$3\sigma$ in the inferred matter density parameter. DESI DR2 BAO data alone yield $\Omega_{\mathrm m}=0.2975\pm0.0086$, lower than the value of approximately $0.315$ preferred by Planck 2018 CMB data~\cite{DESI:2025zgx}. In addition, independent fits to different SN samples, including PantheonPlus, Union3, and DESY5, give central values of $\Omega_{\mathrm m}$ of approximately $0.334$, $0.356$, and $0.352$, respectively, all higher than the CMB-preferred central value~\cite{Brout:2022vxf,DES:2024tys,DESI:2025zgx}. Within the $\Lambda$CDM model, changes in the expansion history induced by variations in neutrino mass can be partially compensated by adjusting other background parameters to preserve the observed CMB acoustic scale, with higher matter densities typically associated with larger neutrino masses. Consequently, the different matter-density preferences of DESI BAO and SN data may affect the joint inference of neutrino mass~\cite{Jiang:2024viw,Du:2024pai}. These results motivate the use of an independent low-redshift probe with observables and sources of systematic uncertainty distinct from those of BAO and SN, to further test the consistency of matter-density inferences and the robustness of neutrino-mass constraints.

Fast radio bursts (FRBs) provide a new low-redshift cosmological probe of the expansion history of the late universe~\cite{Petroff:2021wug,Zhang:2022uzl}. FRBs are intense radio transients, typically lasting milliseconds and predominantly originating outside the Milky Way \cite{Lorimer:2007qn,Thornton:2013iua}, and are commonly classified as repeaters and nonrepeaters \cite{Spitler:2016dmz,Sun:2024huw,Sun:2025cdq,Sun:2026aeh}. Their signals acquire a frequency-dependent arrival-time delay as they propagate through ionized media, which can be characterized by the dispersion measure (DM). For FRBs at cosmological distances, the DM traces the line-of-sight integral of the free-electron number density weighted by cosmological redshift \cite{Deng:2013aga,Zhou:2014yta}. The observed DM mainly includes contributions from the Milky Way medium, the intergalactic medium (IGM), and the host galaxy and local source environment \cite{Dolag:2014bca,Prochaskahalo,Zhang:2020mgq,Zhang:2023pqs,Connor:2024mjg}. Of particular cosmological importance are the contributions from free electrons in the IGM~\cite{McQuinn:2013tmc}; the statistical relation between their mean DM and redshift is commonly called the Macquart relation \cite{Macquart:2020lln}. This relation contains information on the abundance of ionized baryons and also depends on the cosmic expansion history. Therefore, after appropriately accounting for the individual DM contributions and their scatter, localized FRB samples with reliable host-galaxy redshifts can be used to study the expansion history of the late universe \cite{Deng:2013aga,Zhou:2014yta,Gao:2014iva,Walters:2017afr}.

FRBs have broad prospects for cosmological applications \cite{Li:2017mek,Walters:2017afr,Li:2019klc,Wei:2019uhh,Qiang:2020vta,Zhao:2020ole,Li:2020qei,Dai:2021czy,Qiu:2021cww,Hagstotz:2021jzu,Wu:2021jyk,James:2022dcx,Zhao:2022yiv,Yang:2022ftm,Wu:2022dgy,Wang:2022ami,Wei:2023avr,Zhao:2022bpd,Zhang:2023gye,Dai:2023ska,Kalita:2024xae,Piratova-Moreno:2025cpc,Acharya:2025ubt,Gao:2025fcr,Xu:2025ddk,Wu:2026mox,Zhuge:2025urk,Liu:2026epv,Zhang:2025yhi,Zhang:2025rvu,Liu:2025fdf,Wang:2026llr}; for recent reviews of FRB cosmological applications, see Refs. \cite{Zhang:2022uzl,Wu:2024iyu,Wang:2026tgi,Feng:2026cmo}. Recently, localized FRBs have also been combined with DESI DR2 BAO data to simultaneously constrain the Hubble constant ($H_0$) and the cosmic baryon density \cite{Zhang:2026axl}. Previous forecasts showed that future FRB observations have strong potential to constrain $H_0$ and dark-energy parameters. \cite{Zhao:2020ole,Zhang:2023gye,Wu:2022dgy,Zhao:2022bpd,Zhang:2024rra}. For neutrino mass, massive neutrinos alter the mean IGM dispersion measure accumulated along the line of sight over a given redshift interval by affecting the cosmic energy-density composition and expansion history~\cite{Lesgourgues:2006nd}. The DM--redshift relation can therefore provide information on  the expansion of the late universe that complements the CMB in joint neutrino-mass inference, offering a new way to constrain neutrino mass. \citet{Reischke:2023tqh} forecast that, under their adopted model and survey assumptions, combining the DM angular statistics of approximately $5\times10^4$ FRBs with Euclid-like cosmic shear can constrain baryonic feedback and reduce its degeneracy with neutrino mass, improving constraints on the sum of neutrino masses by about a factor of 1.5 relative to cosmic shear alone. \citet{Berghaus:2025kvn} investigated joint neutrino-mass forecasts using FRB DM angular statistics and weak lensing for the Deep Synoptic Array (DSA), and separately examined the inclusion of Planck CMB temperature data and the mean DM--redshift relation. As future radio facilities develop, localized FRB samples with reliable redshifts are expected to grow further, providing more data to investigate the complementarity between FRBs and other observations in different cosmological models and their impact on neutrino-mass inference \cite{Macquart:2015uea,Fialkov:2017qoz,Hashimoto:2020dud}.

In this work, we investigate constraints on the sum of neutrino masses in the $\Lambda$CDM+$\sum m_\nu$ and $w_0w_a$CDM+$\sum m_\nu$ models using CMB, DESI DR2 BAO, DESY5 SN, and 120 localized FRBs. We further forecast the potential of future FRB observations to constrain neutrino masses using mock samples containing 1000 and 10000 events, denoted FRB1(mock) and FRB2(mock), respectively, based on forecasts for the future five-year DSA survey~\cite{Connor:2026nex}.

This paper is organized as follows. Section~\ref{sec2} introduces the cosmological models, the FRB likelihood and mock-sample construction, and the observational data. Section~\ref{sec3} discusses the physical implications of these constraints in the two cosmological models. Section~\ref{sec4} summarizes our conclusions.

\section{METHODOLOGY AND DATA}\label{sec2}

\subsection{Cosmological model}
We first consider the $\Lambda$CDM model with massive neutrinos. In our cosmological parameter inference, we adopt the approximation of three degenerate massive neutrinos, with $m_1=m_2=m_3=\sum m_\nu/3$, and treat the total neutrino mass $\sum m_\nu$ as a free parameter. The total energy density of massive neutrinos is given by \cite{Lesgourgues:2006nd,WMAP:2010qai}
\begin{equation}
\rho_{\nu}(a)=\frac{a^{-4}}{\pi^{2}}\int\frac{q^{2}dq}{e^{q/T_{\nu0}}+1}\sum_{i}\sqrt{q^{2}+m_{i}^{2}a^{2}},
\label{eq:rho_nu}
\end{equation}
where $q$ is the comoving momentum, $m_i$ is the mass of the $i$th neutrino mass eigenstate, and
\begin{equation}
T_{\nu0}=\left(\frac{4}{11}\right)^{1/3}T_{\rm CMB}=1.945\,\mathrm{K},
\end{equation}
is the present neutrino temperature. We take $T_{\rm CMB}=2.725\,\mathrm{K}$. In the early universe, neutrinos are relativistic and
\begin{equation}
\rho_{\nu}(a)\rightarrow N_{\mathrm{eff}}\frac{7}{8}\left(\frac{4}{11}\right)^{4/3}\rho_{\gamma}(a),
\label{eq:rho_nu_early}
\end{equation}
where the effective number of relativistic species is fixed to the Standard Model value $N_{\mathrm{eff}}=3.044$ \cite{Akita:2020szl,Froustey:2020mcq,Bennett:2020zkv}. In the nonrelativistic limit in the late universe,
\begin{equation}
\rho_{\nu}(a)\rightarrow\frac{\sum m_{\nu}}{93.14h^{2}\,\mathrm{eV}}\rho_{\mathrm{crit},0}a^{-3},
\label{eq:rho_nu_late}
\end{equation}
where $\rho_{\mathrm{crit},0}$ is the present critical density.

To investigate the impact of dark-energy dynamics on neutrino-mass constraints, we also consider the $w_0w_a$CDM model. The dimensionless expansion rate $E(z)\equiv H(z)/H_0$ is written as
\begin{equation}
\begin{aligned}
E(z) &= \Bigg[
\Omega_{\mathrm{bc}}(1+z)^3
+\Omega_{\mathrm{de}}f(z) \\
&\quad+\Omega_{\gamma}(1+z)^4
\Bigg(
1+N_{\mathrm{eff}}\frac{7}{8}
\left(\frac{4}{11}\right)^{4/3} \\
&\qquad\qquad\times
\sum_i g\left(
\frac{m_i}{T_{\nu0}(1+z)}
\right)
\Bigg)
\Bigg]^{1/2},
\end{aligned}
\label{eq:Ez}
\end{equation}
where $\Omega_{\mathrm{bc}}=\Omega_{\mathrm{b}}+\Omega_{\mathrm{c}}$
is the sum of the present baryon and cold-dark-matter density
parameters. The quantities $\Omega_{\mathrm{de}}$ and
$\Omega_\gamma$ denote the present dark-energy and photon
density parameters, respectively. The neutrino energy density is included separately through the term containing $g$. Throughout this work, the reported total matter density parameter is defined as $\Omega_{\mathrm{m}}=\Omega_{\mathrm{bc}}+\Omega_\nu$, where $\Omega_\nu$ is the present massive-neutrino density parameter. 
\begin{equation}
g(y)\equiv\frac{40}{7\pi^{4}}\int_{0}^{\infty}\frac{x^{2}\sqrt{x^{2}+y^{2}}}{e^{x}+1}\,dx,
\label{eq:gy}
\end{equation}
and
\begin{equation}
f(z)=(1+z)^{3(1+w_0+w_a)}\exp\left(-\frac{3w_a z}{1+z}\right).
\label{eq:fz}
\end{equation}
The model reduces to $\Lambda$CDM+$\sum m_\nu$ for $w_0=-1$ and $w_a=0$. 

\subsection{Observed FRB dispersion-measure model}\label{sectionIIB}

For the observed FRB sample, we decompose the total observed dispersion measure as
\begin{equation}
{\rm DM}_{\mathrm{obs}} = {\rm DM}_{\mathrm{ISM}} + {\rm DM}_{\mathrm{E}},
\label{eq:DMobs}
\end{equation}
where ${\rm DM}_{\mathrm{ISM}}$ denotes the Milky Way interstellar-medium contribution and ${\rm DM}_{\mathrm{E}}$ is the excess dispersion measure after subtracting this contribution. We estimate ${\rm DM}_{\mathrm{ISM}}$ using the NE2001 Galactic free-electron density model~\cite{Cordes:2002wz}. The excess dispersion measure is further decomposed as
\begin{equation}
{\rm DM}_{\mathrm{E}} = {\rm DM}_{\mathrm{halo}} + {\rm DM}_{\mathrm{IGM}} + \frac{{\rm DM}_{\mathrm{host}}}{1+z}.
\label{eq:DME}
\end{equation}
Here ${\rm DM}_{\mathrm{halo}}$, ${\rm DM}_{\mathrm{IGM}}$, and ${\rm DM}_{\mathrm{host}}$ denote the Milky Way halo, IGM, and host-galaxy contributions, respectively, with ${\rm DM}_{\mathrm{host}}$ defined in the host-galaxy rest frame.

The Milky Way halo contribution is modeled as a Gaussian distribution with mean $\mu_{\mathrm{halo}}=65\,\mathrm{pc\,cm^{-3}}$ and standard deviation $\sigma_{\mathrm{halo}}=15\,\mathrm{pc\,cm^{-3}}$, motivated by the expected Galactic-halo contribution and by FRB analyses adopting the corresponding Gaussian prior \cite{Prochaskahalo,Yang:2022ftm}
\begin{equation}
P_{\mathrm{halo}}({\rm DM}_{\mathrm{halo}})
=
\frac{1}{\sqrt{2\pi}\sigma_{\mathrm{halo}}}
\exp\left[-\frac{({\rm DM}_{\mathrm{halo}}-\mu_{\mathrm{halo}})^2}{2\sigma_{\mathrm{halo}}^2}\right].
\label{eq:Phalo}
\end{equation}

The theoretical mean IGM dispersion measure depends on the expansion history, the baryon density, and the diffuse-baryon fraction $f_{\mathrm d}$, and is given by \cite{Deng:2013aga,Zhou:2014yta,Macquart:2020lln},
\begin{equation}
\left\langle{\rm DM}_{\rm IGM}\right\rangle(z)=
\frac{3cH_0\Omega_{\rm b}f_{\rm d}}{8\pi Gm_{\rm p}}
\int_0^z\frac{(1+z')f_e(z')}{E(z')}\,dz'.
\label{eq:mock_meanDM}
\end{equation}
Here $c$ is the speed of light, $G$ is the gravitational constant, $m_{\mathrm p}$ is the proton mass, and $f_e$ is the free electrons per baryon. Over the redshift range of the observed sample, we assume fully ionized hydrogen and helium with mass fractions of $3/4$ and $1/4$, respectively, and therefore fix $f_e=7/8$. This mean relation is used for both the observed and mock FRB analyses. We model line-of-sight fluctuations in ${\rm DM}_{\mathrm{IGM}}$ with the dimensionless probability distribution \cite{McQuinn:2013tmc,Macquart:2020lln,Zhang:2025wif}
\begin{equation}
P_{\mathrm{IGM}}(\Delta)
=
A\Delta^{-\beta}
\exp\left[-\frac{(\Delta^{-\alpha}-C_0)^2}{2\alpha^2\sigma_{\mathrm{IGM}}^2}\right],
\qquad \Delta>0,
\label{eq:Pigm}
\end{equation}
where
\begin{equation}
\Delta\equiv\frac{{\rm DM}_{\mathrm{IGM}}}{\langle{\rm DM}_{\mathrm{IGM}}\rangle}.
\end{equation}
We fix $\alpha=3$ and $\beta=3$ and parameterize the effective scatter as
\begin{equation}
\sigma_{\mathrm{IGM}}=Fz^{-1/2}.
\end{equation}
Here $F$ is a free dimensionless feedback parameter, $\sigma_{\mathrm{IGM}}$ is also dimensionless, $A$ is a normalization constant, and $C_0$ is chosen so that the dimensionless distribution has unit mean \cite{Macquart:2020lln}.

The host-galaxy rest-frame contribution is modeled as a log-normal distribution \cite{Macquart:2020lln},
\begin{equation}
\begin{aligned}
P_{\mathrm{host}}({\rm DM}_{\mathrm{host}})
&=
\frac{1}{\sqrt{2\pi}\sigma_{\mathrm{host}}{\rm DM}_{\mathrm{host}}} \\
&\quad\times
\exp\left[-\frac{(\ln {\rm DM}_{\mathrm{host}}-\mu_{\mathrm{host}})^2}{2\sigma_{\mathrm{host}}^2}\right].
\end{aligned}
\label{eq:Phost}
\end{equation}
The sampled parameter $e^{\mu_{\mathrm{host}}}$ is the median of the host-galaxy rest-frame DM distribution, whereas $\sigma_{\mathrm{host}}$ is its dimensionless logarithmic width.

For each observed FRB, the theoretical probability $P_i$ is obtained by convolving the probability distributions of the Milky Way halo, IGM, and host-galaxy contributions. The corresponding FRB log-likelihood is
\begin{equation}
\ln\mathcal{L}_{\mathrm{FRB}}
=
\sum_{i=1}^{N_{\mathrm{FRB}}}\ln P_i.
\label{eq:Likelihood}
\end{equation}

\subsection{Mock FRB catalogs}\label{sectionIIC}

The mock catalogs are constructed using a $\Lambda$CDM fiducial cosmology, with the mean IGM dispersion measure calculated from Eq.~\eqref{eq:mock_meanDM}. We adopt $\Omega_{\mathrm{m}}=0.3153$, $h=0.6736$, $\Omega_{\mathrm b}h^2=0.02237$, $N_{\mathrm{eff}}=3.044$, $T_{\mathrm{CMB}}=2.7255\,\mathrm K$, and $\sum m_\nu=0.06\,\mathrm{eV}$. We fix $f_{\mathrm d}=0.90$ and $f_e=7/8$ over the simulated redshift range.

Redshifts are drawn over $0\leq z\leq2.5$ from the constant-comoving-density form with a Gaussian detection cutoff introduced in Ref.~\cite{Munoz:2016tmg},
\begin{equation}
p(z)\propto \frac{D_{\rm C}^2(z)}{H(z)(1+z)}
\exp\left[-\frac{D_{\rm L}^2(z)}{2D_{\rm L}^2(1)}\right],
\label{eq:mock_pz}
\end{equation}
where $D_{\rm C}$ and $D_{\rm L}$ are the comoving and luminosity distances, respectively.

The total observer-frame DM uncertainty assigned to each mock event is
\begin{equation}
\begin{aligned}
\sigma_{\mathrm{DM}}^{\mathrm{mock}}(z)=\Big[
&\sigma_{\mathrm{obs}}^2+\sigma_{\mathrm{MW}}^2+\sigma_{\mathrm{halo}}^2\\
&+\left(\sigma_{\mathrm{IGM}}^{\mathrm{mock}}(z)\right)^2
+\left(\frac{\sigma_{\mathrm{host}}^{\mathrm{mock}}}{1+z}\right)^2
\Big]^{1/2},
\end{aligned}
\label{eq:mock_sigmaDM}
\end{equation}
with
\begin{equation}
\sigma_{\mathrm{IGM}}^{\mathrm{mock}}(z)
=173.8~z^{0.4}\,\mathrm{pc\,cm^{-3}},
\label{eq:mock_sigmaIGM}
\end{equation}
following the empirical scatter prescription adopted in recent FRB forecasts \cite{Qiang:2020vta}.
In the observer frame, we fix $\sigma_{\mathrm{obs}}=1.5$, $\sigma_{\mathrm{MW}}=10$, $\sigma_{\mathrm{halo}}=15$, and $\sigma_{\mathrm{host}}^{\mathrm{mock}}=30$, all in units of $\mathrm{pc\,cm^{-3}}$. The mean host dispersion measure is fixed to $50/(1+z)\,\mathrm{pc\,cm^{-3}}$ in the observer frame.

The simulated dispersion measure is generated as
\begin{equation}
\begin{aligned}
{\rm DM}_{\mathrm{sim}}=\max\Bigg[0,\,
&\left\langle{\rm DM}_{\mathrm{IGM}}\right\rangle(z)
+\frac{50\,\mathrm{pc\,cm^{-3}}}{1+z}\\
&+\sigma_{\mathrm{DM}}^{\mathrm{mock}}(z)\epsilon
\Bigg],\qquad\epsilon\sim\mathcal N(0,1),
\end{aligned}
\label{eq:mock_DM}
\end{equation}
with negative draws set to zero. The generated mock catalogs contain the redshift, simulated dispersion measure, assigned DM uncertainty, and corresponding mean IGM dispersion measure for each event.

For cosmological parameter inference with the mock FRB samples, we adopt a Gaussian likelihood based on the mean DM--redshift relation. The theoretical prediction is
\begin{equation}
{\rm DM}_{\mathrm{th}}(z)
=\left\langle{\rm DM}_{\mathrm{IGM}}\right\rangle(z)
+\frac{50\,\mathrm{pc\,cm^{-3}}}{1+z}.
\label{eq:mock_dmth}
\end{equation}
The mean IGM dispersion measure is calculated by numerically integrating Eq.~\eqref{eq:mock_meanDM}, with $H(z)$ provided by \texttt{CAMB} \cite{Lewis:1999bs,Howlett:2012mh}. The mean host contribution is the same as that adopted in the mock generation. For the $i$th event, the uncertainty $\sigma_{\mathrm{DM},i}^{\mathrm{mock}}$ is evaluated from Eq.~\eqref{eq:mock_sigmaDM} at $z_i$.

Assuming independent Gaussian DM errors for different events, the mock-sample log-likelihood is \cite{Walters:2017afr,Jaroszynski:2018vgh}
\begin{equation}
\ln\mathcal L_{\mathrm{FRB,mock}}
=-\frac12\sum_{i=1}^{N_{\mathrm{FRB}}}
\frac{\left[{\rm DM}_{\mathrm{sim},i}-{\rm DM}_{\mathrm{th}}(z_i)\right]^2}
{\left(\sigma_{\mathrm{DM},i}^{\mathrm{mock}}\right)^2}.
\label{eq:mock_likelihood}
\end{equation}
Here ${\rm DM}_{\mathrm{sim},i}$ and $z_i$ are the simulated dispersion measure and redshift of the $i$th event, respectively, and ${\rm DM}_{\mathrm{th}}(z_i)$ is the corresponding theoretical prediction. The mock uncertainty prescription is fixed.

We emphasize that different likelihood functions are employed for the observed and mock FRB samples. For the observed FRBs, the likelihood is constructed by convolving the probability distributions of the Milky Way halo, IGM, and host-galaxy contributions, as described in Section~\ref{sectionIIB} and Eq.~\eqref{eq:Likelihood}. This treatment accounts for non-Gaussian IGM fluctuations and the associated astrophysical uncertainties. In contrast, the mock FRB analysis adopts a Gaussian likelihood based on the mean DM--redshift relation and a prescribed uncertainty model to assess the potential constraining power of future large FRB samples. 

The five-year DSA/Chronoscope survey is expected to detect  $\sim 10^5$ FRBs with localization precision of approximately 250 milliarcseconds~\cite{Connor:2026nex}. Assuming that $10\%$ of the detected events can be precisely localized, securely associated with their host galaxies, and have their redshifts measured, the survey would yield $\sim 10^4$ localized FRBs with redshift information for cosmological parameter inference. In this work, we consider two representative scenarios: a conservative case with $N_{\mathrm{FRB}}=1000$ and a normal case with $N_{\mathrm{FRB}}=10000$.

\subsection{Data and parameter inference}

\textbf{FRB:} We adopt the 120-event observed FRB sample compiled by Ref.~\cite{Zhang:2026axl}. This sample was selected from an initial compilation of 139 FRBs by applying the following criteria: (i) a valid redshift satisfying $z>10^{-5}$; (ii)  ${\rm DM}_{\rm ISM}$ no larger than $40\%$ of the observed dispersion measure ${\rm DM}_{\rm obs}$ \cite{Connor:2024mjg}; (iii) an effective extragalactic dispersion measure ${\rm DM}_{\rm E}>80\,\mathrm{pc\,cm^{-3}}$; and (iv) the exclusion of FRB 20190520B and FRB 20220831A owing to their unusually large local dispersion contributions \cite{Connor:2024mjg}. We denote this sample by FRB.

\textbf{FRB1(mock):} Following the mock-generation procedure described in Section~\ref{sectionIIC}, we construct a mock sample containing 1000 FRBs with redshift information and denote it by FRB1(mock).

\textbf{FRB2(mock):} Following the same procedure, we construct a mock sample containing 10000 FRBs with redshift information and denote it by FRB2(mock).

\textbf{CMB:} The CMB likelihood consists of four components: (i) the Planck \texttt{CamSpec} small-scale ($\ell>30$) TT, TE, and EE spectra \cite{Efstathiou:2019mdh,Rosenberg:2022sdy}; (ii) the Planck \texttt{Commander} large-scale ($2\leq\ell\leq30$) TT spectrum; (iii) the Planck \texttt{SimAll} large-scale EE likelihood \cite{Planck:2018vyg,Planck:2019nip}; and (iv) CMB lensing using the PR4 reconstruction \cite{Carron:2022eyg} together with Atacama Cosmology Telescope Data Release 6 (ACT DR6) lensing \cite{ACT:2023dou}.

\textbf{DESI:} We use DESI DR2 BAO measurements from galaxies, quasars, and the Ly$\alpha$ forest \cite{DESI:2025zgx}.

\textbf{DESY5:} The DESY5 sample contains 1829 Type Ia supernovae \cite{DES:2024tys}: 1635 photometrically classified DES supernovae at $0.10<z<1.13$ and 194 spectroscopically confirmed low-redshift supernovae at $0.025\leq z\leq0.10$.

We perform Markov Chain Monte Carlo (MCMC) parameter inference \cite{Lewis:2002ah,Lewis:2013hha} with the publicly available Bayesian inference framework \texttt{Cobaya}~\cite{Torrado:2020dgo}, using \texttt{CAMB} \cite{Lewis:1999bs,Howlett:2012mh} to compute the cosmological background evolution and CMB observables. Chain convergence is assessed with the Gelman--Rubin criterion $R-1<0.02$ \cite{Gelman:1992zz}, and marginalized posteriors are analyzed with \texttt{GetDist} \cite{Lewis:2019xzd}. The flat priors of sampling parameters are summarized in Table~\ref{tab:priors}.

\begin{table}[htbp]
\centering
\caption{Flat priors on the main cosmological and FRB parameters.}
\label{tab:priors}
\renewcommand{\arraystretch}{1.35}
\begin{tabular}{ccc}
\hline\hline
\textbf{Model/Data} & \textbf{Parameter} & \textbf{Prior} \\
\hline
& $\Omega_\mathrm{b}h^2$ & $\mathcal{U}[0.005,0.1]$ \\
& $\Omega_\mathrm{c}h^2$ & $\mathcal{U}[0.001,0.99]$ \\
& $\tau_\mathrm{reio}$ & $\mathcal{U}[0.01,0.8]$ \\
$\Lambda$CDM & $\log(10^{10}A_\mathrm{s})$ & $\mathcal{U}[1.61,3.91]$ \\
& $n_\mathrm{s}$ & $\mathcal{U}[0.8,1.2]$ \\
& $H_0~[{\rm km\,s^{-1}\,Mpc^{-1}}]$ & $\mathcal{U}[20,100]$ \\
& $\sum m_\nu\,[\mathrm{eV}]$ & $\mathcal{U}[0,5]$ \\
\hline
\multirow[c]{2}{*}{$w_0w_a$CDM}
& $w_0$ & $\mathcal{U}[-3,1]$ \\
& $w_a$ & $\mathcal{U}[-3,2]$ \\
\hline
\multirow[c]{4}{*}{FRB}
& $f_\mathrm{d}$ & $\mathcal{U}[0.84,0.96]$ \\
& $e^{\mu_\mathrm{host}}~[\rm pc ~cm^{-3}]$ & $\mathcal{U}[20,200]$ \\
& $\sigma_\mathrm{host}$ & $\mathcal{U}[0.2,2.0]$ \\
& $F$ & $\mathcal{U}[0.011,0.5]$ \\
\hline\hline
\end{tabular}
\end{table}

\begin{table*}[t]
\centering
\setlength{\tabcolsep}{6pt}
\caption{Marginalized constraints from the observed-data combinations in the $\Lambda$CDM+$\sum m_\nu$ and $w_0w_a$CDM+$\sum m_\nu$ models. Uncertainties are given at the $68\%$ confidence level (CL), while upper limits on $\sum m_{\nu}$ are given at the $95\%$ CL. }
\label{tab:results}
\renewcommand{\arraystretch}{1.25}
\begin{tabular*}{\textwidth}{@{\extracolsep{\fill}}lccccc}
\hline\hline
\textbf{Model/Dataset} & $\Omega_\mathrm{m}$ & $H_0\,[\mathrm{km\,s^{-1}\,Mpc^{-1}}]$ & $\sum m_\nu\,[\mathrm{eV}]$ & $w_0$ & $w_a$ \\
\hline
\textbf{$\bm{\Lambda}$CDM+$\bm{\sum m_\nu}$} & & & & & \\
CMB+DESI & $0.3009\pm0.0036$ & $68.37\pm0.29$ & $<0.068$ & $-$ & $-$ \\
CMB+FRB & $0.3180^{+0.0076}_{-0.0130}$ & $67.10^{+1.00}_{-0.60}$ & $<0.21$ & $-$ & $-$ \\
CMB+DESI+DESY5 & $0.3034^{+0.0033}_{-0.0037}$ & $68.18\pm0.28$ & $<0.077$ & $-$ & $-$ \\
CMB+DESI+FRB & $0.3006\pm0.0037$ & $68.39\pm0.29$ & $<0.068$ & $-$ & $-$ \\
CMB+DESI+FRB+DESY5 & $0.3035\pm0.0037$ & $68.17\pm0.29$ & $<0.079$ & $-$ & $-$ \\
\hline
\textbf{$\bm{w_0w_a}$CDM+$\bm{\sum m_\nu}$} & & & & & \\
CMB+DESI & $0.3530\pm0.0220$ & $63.70^{+1.70}_{-2.20}$ & $<0.17$ & $-0.42^{+0.23}_{-0.21}$ & $-1.75\pm0.62$ \\
CMB+FRB & $0.1799^{+0.0086}_{-0.0370}$ & $>87.80$ & $<0.29$ & $-1.63^{+0.33}_{-0.47}$ & $-$ \\
CMB+DESI+DESY5 & $0.3188\pm0.0058$ & $66.76\pm0.57$ & $<0.13$ & $-0.76\pm0.06$ & $-0.83^{+0.26}_{-0.21}$ \\
CMB+DESI+FRB & $0.3430\pm0.0220$ & $64.50^{+1.80}_{-2.10}$ & $<0.15$ & $-0.51\pm0.21$ & $-1.50^{+0.68}_{-0.59}$ \\
CMB+DESI+FRB+DESY5 & $0.3181\pm0.0058$ & $66.82\pm0.57$ & $<0.13$ & $-0.77\pm0.06$ & $-0.80^{+0.26}_{-0.20}$ \\
\hline\hline
\end{tabular*}
\end{table*}

\begin{figure*}[t]
\centering
\subfigure{%
\includegraphics[width=0.47\textwidth]{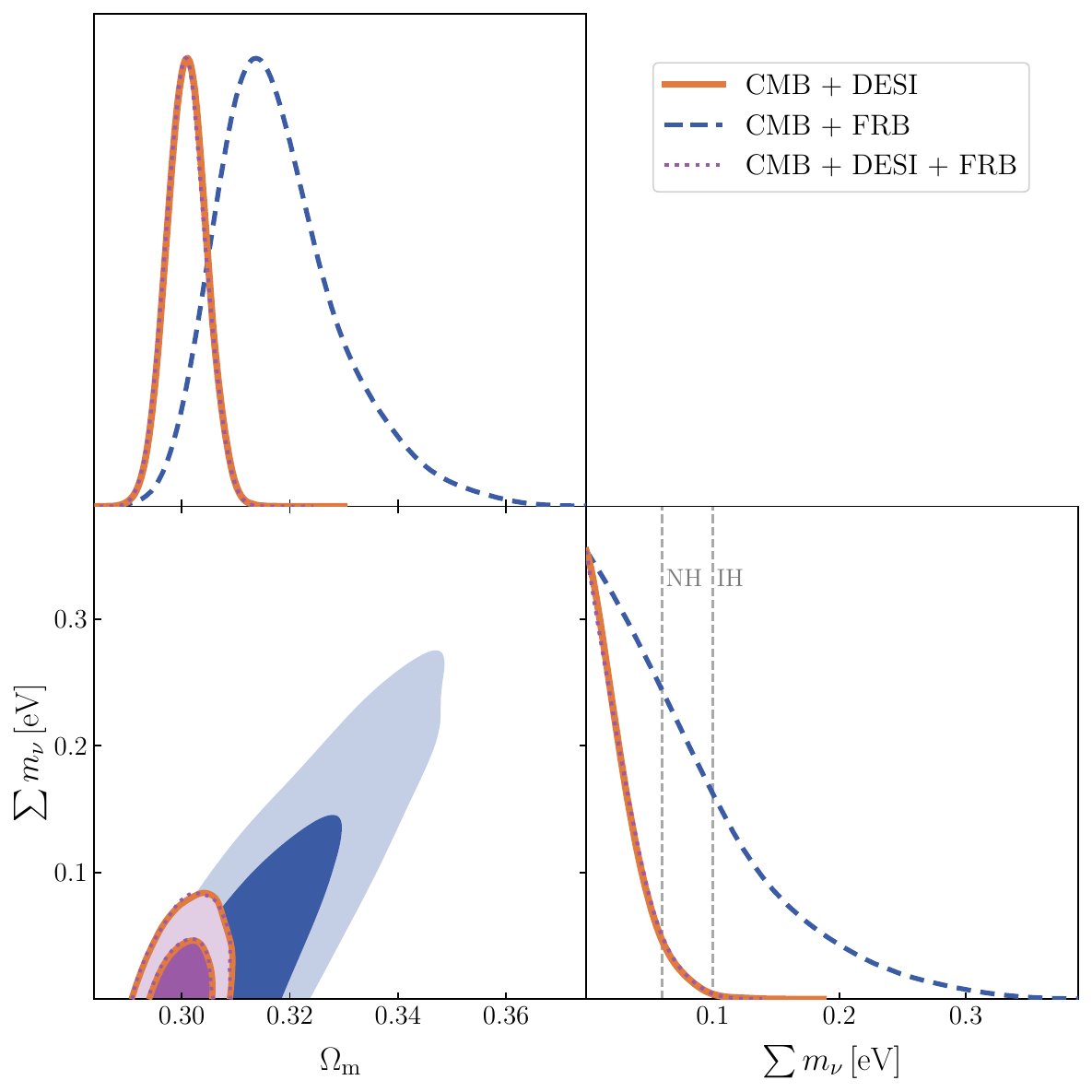}}
\hfill
\subfigure{%
\includegraphics[width=0.47\textwidth]{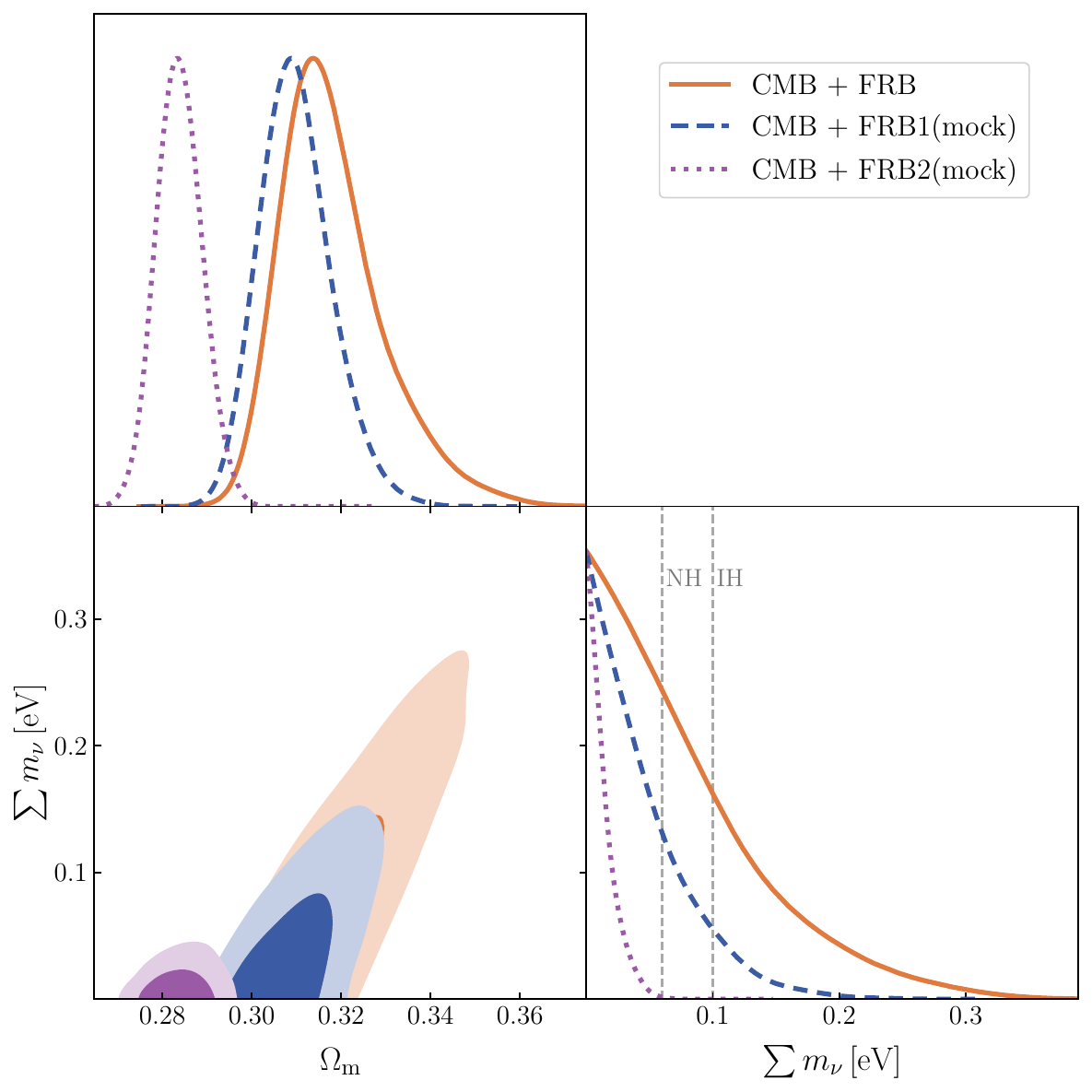}}
\caption{Marginalized 1D and 2D posterior distributions in the $\Lambda$CDM+$\sum m_\nu$ model, with 68\% and 95\% contours. Gray vertical dashed lines indicate the minimum neutrino-mass sums for NH and IH. \textbf{\textit{Left panel:}} Constraints from current observational data. \textbf{\textit{Right panel:}} Comparison with mock FRB samples. FRB, FRB1(mock), and FRB2(mock) contain 120 observed, 1000 mock, and 10000 mock events, respectively.}
\label{fig:lambda_pair}
\end{figure*}

\begin{table*}[t]
\centering
\setlength{\tabcolsep}{8pt}
\caption{Constraints from CMB combined with the mock FRB catalogs in the $\Lambda$CDM+$\sum m_\nu$ and $w_0w_a$CDM+$\sum m_\nu$ models. Uncertainties are given at the $68\%$ CL, while upper limits on $\sum m_{\nu}$ are given at the $95\%$ CL.}
\label{tab:simulation}
\renewcommand{\arraystretch}{1.25}
\begin{tabular}{lccccc}
\hline\hline
\textbf{Model/Dataset} & $\Omega_\mathrm{m}$ & $H_0\,[\mathrm{km\,s^{-1}\,Mpc^{-1}}]$ & $\sum m_\nu\,[\mathrm{eV}]$ & $w_0$ & $w_a$ \\
\hline
\textbf{$\bm{\Lambda}$CDM+$\bm{\sum m_\nu}$} & & & & & \\
CMB+FRB1(mock) & $0.3097^{+0.0070}_{-0.0090}$ & $67.70^{+0.69}_{-0.54}$ & $<0.12$ & $-$ & $-$ \\
CMB+FRB2(mock) & $0.2838\pm0.0052$ & $69.74\pm0.44$ & $<0.035$ & $-$ & $-$ \\
\hline
\textbf{$\bm{w_0w_a}$CDM+$\bm{\sum m_\nu}$} & & & & & \\
CMB+FRB1(mock) & $0.2020^{+0.0240}_{-0.0410}$ & $85.10\pm6.80$ & $<0.30$ & $-1.60^{+0.34}_{-0.55}$ & $-0.02^{+1.80}_{-0.69}$ \\
CMB+FRB2(mock) & $0.1848^{+0.0088}_{-0.0130}$ & $88.60^{+2.90}_{-2.60}$ & $<0.38$ & $-2.06^{+0.13}_{-0.20}$ & $>1.29$ \\
\hline\hline
\end{tabular}
\end{table*}

\begin{figure*}[htb]
\centering
\includegraphics[width=0.92\textwidth]{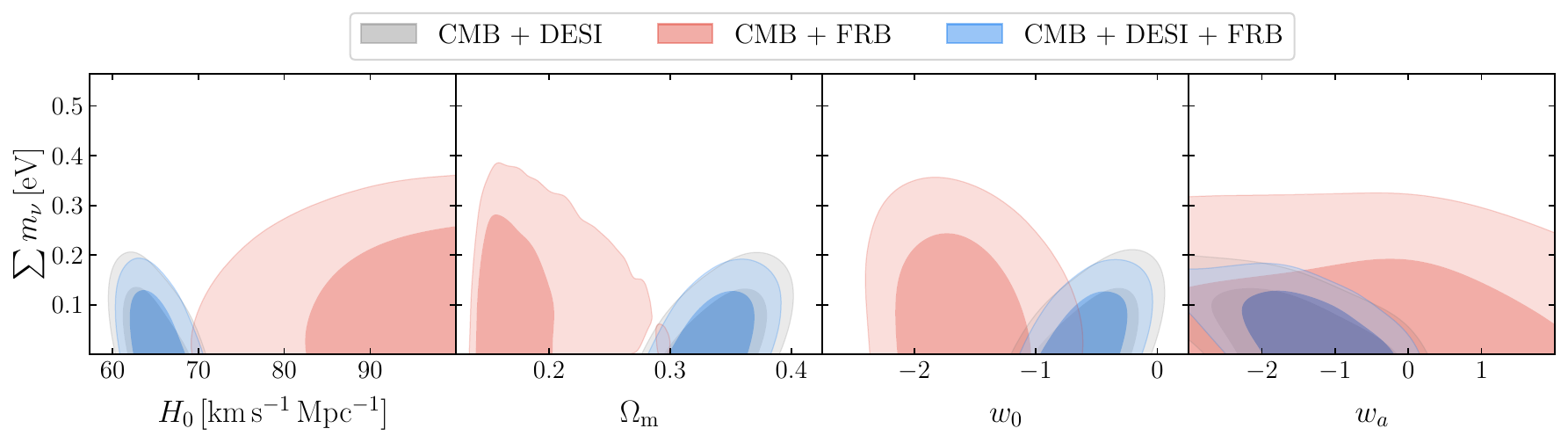}
\caption{Marginalized two-dimensional posterior distributions of $\sum m_\nu$ with $H_0$, $\Omega_{\rm m}$, $w_0$, and $w_a$ in the $w_0w_a$CDM+$\sum m_\nu$ model for CMB+DESI, CMB+FRB, and CMB+DESI+FRB.}
\label{fig:w0wa_current}
\end{figure*}
\begin{figure*}[!htbp]
\centering
\includegraphics[width=0.92\textwidth]{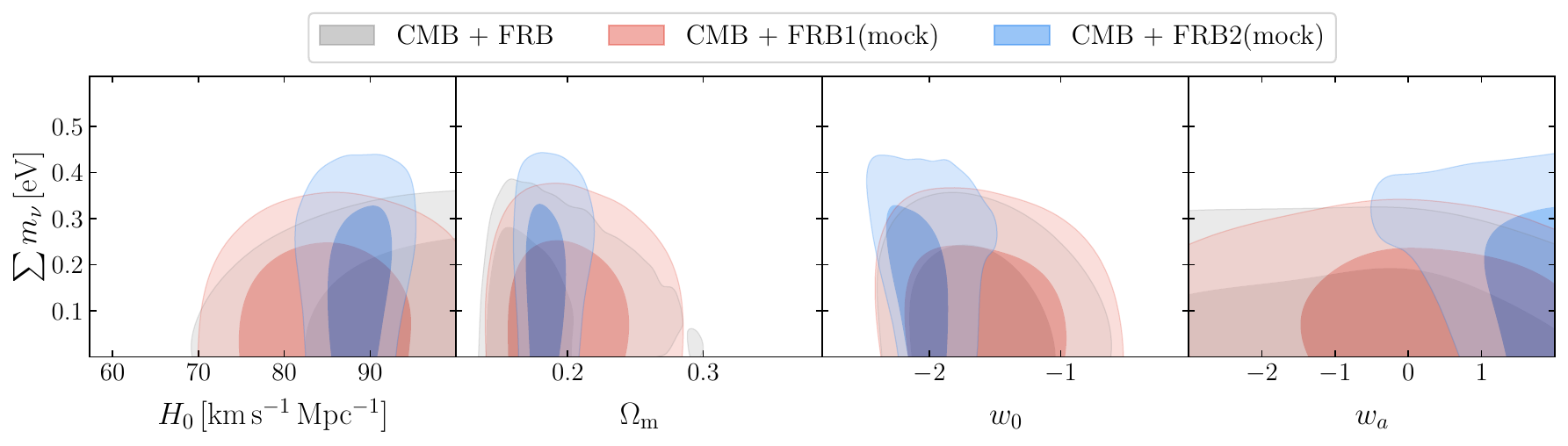}
\caption{Marginalized two-dimensional posterior distributions in the $w_0w_a$CDM+$\sum m_\nu$ model for CMB+FRB, CMB+FRB1(mock), and CMB+FRB2(mock).}
\label{fig:w0wa_mock}
\end{figure*}

\section{RESULTS AND DISCUSSION}\label{sec3}

In this section, we present the parameter constraints for the $\Lambda$CDM+$\sum m_\nu$ and $w_0w_a$CDM+$\sum m_\nu$ models in Table~\ref{tab:results}, while Table~\ref{tab:simulation} gives the constraints from CMB combined with FRB1(mock) and FRB2(mock). The one-dimensional marginalized posterior distributions and two-dimensional contours of the main parameters are shown in Figs.~\ref{fig:lambda_pair}--\ref{fig:w0wa_mock}.

In the $\Lambda$CDM+$\sum m_\nu$ model, CMB+FRB gives $\sum m_\nu<0.21\,\mathrm{eV}$, substantially relaxed compared with the CMB+DESI limit of $0.068\,\mathrm{eV}$. This difference is related to the different matter-density preferences of the two data combinations: CMB+FRB gives $\Omega_{\rm m}=0.3180^{+0.0076}_{-0.0130}$, higher than the CMB+DESI value of $\Omega_{\rm m}=0.3009\pm0.0036$, as shown in the left panel of Fig.~\ref{fig:lambda_pair}. At the background level, nonrelativistic neutrinos contribute to the matter density in the later universe. Under the CMB constraints on the physical matter densities in the early universe and the acoustic scale, higher $\Omega_{\rm m}$ is associated with lower $H_0$ and larger neutrino masses~\cite{Lesgourgues:2006nd}. Consequently, the higher-$\Omega_{\rm m}$ region allowed by CMB+FRB admits larger neutrino masses. At the perturbation level, the enhancement of structure growth and lensing power associated with a higher matter density can partially compensate for the suppression of small-scale power induced by the free streaming of more massive neutrinos, thereby also relaxing the neutrino-mass constraint.

Further joint analyses of FRB with CMB, DESI, and DESY5 show a very small contribution from FRB, indicating that the additional statistical contribution of the current FRB sample is limited relative to the other datasets in this combination. To investigate the constraining potential of larger samples, we construct larger mock FRB samples and analyze them jointly with the CMB. Under the mock assumptions adopted in this work, when the sample reaches 1000 events, CMB+FRB1(mock) gives $\sum m_\nu<0.12\,\mathrm{eV}$, approximately 43\% lower than the current CMB+FRB limit of $0.21\,\mathrm{eV}$, but still above the CMB+DESI limit of $0.068\,\mathrm{eV}$. When the sample is further increased to 10000 events, the CMB+FRB2(mock) limit decreases to $0.035\,\mathrm{eV}$, approximately 83\% lower than the current CMB+FRB limit and more stringent than the CMB+DESI neutrino-mass limit, as shown in Table~\ref{tab:simulation}. The precision of other cosmological parameters also increases substantially, indicating that neutrino-mass constraints from future localized FRB samples containing 10000 events will surpass those provided by current DESI data.

When the cosmological model is extended from $\Lambda$CDM to $w_0w_a$CDM, the upper bound on the sum of neutrino masses is substantially relaxed, demonstrating a significant cosmological-model dependence of neutrino-mass constraints. In the $w_0w_a$CDM+$\sum m_\nu$ model, CMB+FRB gives $\sum m_\nu<0.29\,\mathrm{eV}$, relaxed relative to the $0.21\,\mathrm{eV}$ limit in $\Lambda$CDM+$\sum m_\nu$. This combination also yields $H_0>87.80\,\mathrm{km\,s^{-1}\,Mpc^{-1}}$, $\Omega_{\rm m}=0.1799^{+0.0086}_{-0.0370}$, and $w_0=-1.63^{+0.33}_{-0.47}$, while $w_a$ is not effectively constrained, as shown in Fig.~\ref{fig:w0wa_current} and Table~\ref{tab:results}. At the background level, $w_0<-1$ indicates that dark energy is currently in the phantom regime. At the same dark-energy density, its more negative pressure makes a stronger contribution to cosmic acceleration than $\Lambda$CDM, yielding a higher inferred $H_0$. Meanwhile, the negative correlation between $\Omega_{\rm m}$ and $H_0$ means that higher $H_0$ is accompanied by lower $\Omega_{\rm m}$. In addition, CMB+DESI gives a neutrino-mass limit of $0.17\,\mathrm{eV}$, which becomes $0.15\,\mathrm{eV}$ after adding FRB, while the CMB+DESI+DESY5 limit changes little when FRB is added. As in the $\Lambda$CDM model, this indicates that the FRB sample at its current size provides only limited additional constraining power when combined with other cosmological probes.

Because the current observed FRB sample is limited in size, we further use mock data to investigate the constraints obtained with larger samples. In the $w_0w_a$CDM+$\sum m_\nu$ model, when the sample reaches 1000 events, CMB+FRB1(mock) gives $\sum m_\nu<0.30\,\mathrm{eV}$, above the $0.12\,\mathrm{eV}$ limit from the same mock sample in $\Lambda$CDM+$\sum m_\nu$. When the sample is further increased to 10000 events, the CMB+FRB2(mock) neutrino-mass  limit is relaxed to $0.38\,\mathrm{eV}$, also above the $0.035\,\mathrm{eV}$ limit from the same sample in $\Lambda$CDM+$\sum m_\nu$. As shown in Fig.~\ref{fig:w0wa_mock} and Table~\ref{tab:simulation}, FRB2(mock) further narrows the posterior ranges of $\Omega_{\rm m}$, $H_0$, and $w_0$ relative to FRB1(mock), while shifting the posterior locations toward lower $\Omega_{\rm m}$, higher $H_0$, and more negative $w_0$. Because a joint degeneracy between neutrino mass and dark-energy evolution remains, these changes in the posterior locations of other parameters also affect the marginalized neutrino-mass limit. 

\section{CONCLUSION}\label{sec4}

In this work, we investigate the ability of the FRB DM--redshift relation, combined with the CMB, to constrain neutrino mass and compare FRBs with DESI BAO and SN data as low-redshift probes, focusing on the effects of FRB sample size and the dark-energy model in the later universe on the inferred results. 

In the $\Lambda$CDM+$\sum m_\nu$ model, the CMB+FRB analysis based on 120 observed localized FRBs gives $\sum m_\nu<0.21\,\mathrm{eV}$, substantially relaxed relative to joint constraints from current CMB, DESI, and DESY5 data. To forecast the neutrino-mass constraints from larger future FRB datasets, we find that the CMB+FRB1(mock) limit decreases to $0.12\,\mathrm{eV}$ when the sample reaches 1000 events; increasing the sample to 10000 events further decreases the CMB+FRB2(mock) limit to $0.035\,\mathrm{eV}$. This already provides a stronger neutrino-mass constraint than current CMB+DESI, demonstrating the potential of large localized FRB samples to constrain neutrino mass. When the model is extended to $w_0w_a$CDM+$\sum m_\nu$, the CMB+FRB neutrino-mass limit is relaxed to $0.29\,\mathrm{eV}$. For 1000 simulated FRBs, CMB+FRB1(mock) gives a limit of $0.30\,\mathrm{eV}$; increasing the sample to 10000 events gives a CMB+FRB2(mock) limit of $0.38\,\mathrm{eV}$. These limits are all higher than the corresponding results from the same samples in $\Lambda$CDM+$\sum m_\nu$, demonstrating a clear model dependence of neutrino-mass inference and showing that the constraints from larger samples remain affected by parameter degeneracies.

Overall, our results highlight the potential of localized FRBs for neutrino-mass inference, as well as the importance of the assumed dark-energy model. As larger FRB samples with reliable host-galaxy redshifts become available and the modeling of the IGM and host-galaxy DM contributions improves, FRBs may provide more precise independent constraints on the late-time expansion history, improve neutrino-mass constraints, and help assess their robustness across different cosmological models.

\section*{Acknowledgments}
We thank Wan-Peng Sun, Sheng-Han Zhou, and Yi-Min Zhang for their fruitful discussions. This work was supported by the National Natural Science Foundation of China (Grants Nos.~12533001, 12575049, 12473001), the National SKA Program of China (Grants Nos.~2022SKA0110200 and 2022SKA0110203), the China Manned Space Program (Grant No.~CMS-CSST-2025-A02), and the 111 Project (Grant No.~B16009).

\bibliography{main}

@article{Super-Kamiokande:1998kpq,
    author = "Fukuda, Y. and others",
    collaboration = "Super-Kamiokande",
    title = "{Evidence for oscillation of atmospheric neutrinos}",
    eprint = "hep-ex/9807003",
    archivePrefix = "arXiv",
    reportNumber = "BU-98-17, ICRR-REPORT-422-98-18, UCI-98-8, KEK-PREPRINT-98-95, LSU-HEPA-5-98, UMD-98-003, SBHEP-98-5, TKU-PAP-98-06, TIT-HPE-98-09",
    doi = "10.1103/PhysRevLett.81.1562",
    journal = "Phys. Rev. Lett.",
    volume = "81",
    pages = "1562--1567",
    year = "1998"
}

@article{SNO:2002tuh,
    author = "Ahmad, Q. R. and others",
    collaboration = "SNO",
    title = "{Direct evidence for neutrino flavor transformation from neutral current interactions in the Sudbury Neutrino Observatory}",
    eprint = "nucl-ex/0204008",
    archivePrefix = "arXiv",
    doi = "10.1103/PhysRevLett.89.011301",
    journal = "Phys. Rev. Lett.",
    volume = "89",
    pages = "011301",
    year = "2002"
}

@article{ParticleDataGroup:2024cfk,
    author = "Navas, S. and others",
    collaboration = "Particle Data Group",
    title = "{Review of particle physics}",
    doi = "10.1103/PhysRevD.110.030001",
    journal = "Phys. Rev. D",
    volume = "110",
    number = "3",
    pages = "030001",
    year = "2024"
}

@article{Esteban:2020cvm,
    author = "Esteban, Ivan and Gonzalez-Garcia, M. C. and Maltoni, Michele and Schwetz, Thomas and Zhou, Albert",
    title = "{The fate of hints: updated global analysis of three-flavor neutrino oscillations}",
    eprint = "2007.14792",
    archivePrefix = "arXiv",
    primaryClass = "hep-ph",
    reportNumber = "IFT-UAM/CSIC-112, YITP-SB-2020-21",
    doi = "10.1007/JHEP09(2020)178",
    journal = "JHEP",
    volume = "09",
    pages = "178",
    year = "2020"
}

@article{deSalas:2020pgw,
    author = "de Salas, P. F. and Forero, D. V. and Gariazzo, S. and Mart{\'\i}nez-Mirav{\'e}, P. and Mena, O. and Ternes, C. A. and T{\'o}rtola, M. and Valle, J. W. F.",
    title = "{2020 global reassessment of the neutrino oscillation picture}",
    eprint = "2006.11237",
    archivePrefix = "arXiv",
    primaryClass = "hep-ph",
    doi = "10.1007/JHEP02(2021)071",
    journal = "JHEP",
    volume = "02",
    pages = "071",
    year = "2021"
}

@article{KATRIN:2024cdt,
    author = "Aker, Max and others",
    collaboration = "KATRIN",
    title = "{Direct neutrino-mass measurement based on 259 days of KATRIN data}",
    eprint = "2406.13516",
    archivePrefix = "arXiv",
    primaryClass = "nucl-ex",
    doi = "10.1126/science.adq9592",
    journal = "Science",
    volume = "388",
    number = "6743",
    pages = "adq9592",
    year = "2025"
}

@article{Lesgourgues:2006nd,
    author = "Lesgourgues, Julien and Pastor, Sergio",
    title = "{Massive neutrinos and cosmology}",
    eprint = "astro-ph/0603494",
    archivePrefix = "arXiv",
    reportNumber = "LAPTH-1131-05, IFIC-05-59",
    doi = "10.1016/j.physrep.2006.04.001",
    journal = "Phys. Rept.",
    volume = "429",
    pages = "307--379",
    year = "2006"
}

@article{Zhang:2014nta,
    author = "Zhang, Jing-Fei and Li, Yun-He and Zhang, Xin",
    title = "{Cosmological constraints on neutrinos after BICEP2}",
    eprint = "1404.3598",
    archivePrefix = "arXiv",
    primaryClass = "astro-ph.CO",
    doi = "10.1140/epjc/s10052-014-2954-8",
    journal = "Eur. Phys. J. C",
    volume = "74",
    pages = "2954",
    year = "2014"
}

@article{TopicalConvenersKNAbazajianJECarlstromATLee:2013bxd,
    author = "Abazajian, K. N. and others",
    collaboration = "Topical Conveners: K.N. Abazajian, J.E. Carlstrom, A.T. Lee",
    title = "{Neutrino Physics from the Cosmic Microwave Background and Large Scale Structure}",
    eprint = "1309.5383",
    archivePrefix = "arXiv",
    primaryClass = "astro-ph.CO",
    reportNumber = "FERMILAB-PUB-13-438-A",
    doi = "10.1016/j.astropartphys.2014.05.014",
    journal = "Astropart. Phys.",
    volume = "63",
    pages = "66--80",
    year = "2015"
}

@article{Zhang:2014dxk,
    author = "Zhang, Jing-Fei and Li, Yun-He and Zhang, Xin",
    title = "{Sterile neutrinos help reconcile the observational results of primordial gravitational waves from Planck and BICEP2}",
    eprint = "1403.7028",
    archivePrefix = "arXiv",
    primaryClass = "astro-ph.CO",
    doi = "10.1016/j.physletb.2014.12.012",
    journal = "Phys. Lett. B",
    volume = "740",
    pages = "359--363",
    year = "2015"
}

@article{Planck:2018vyg,
    author = "Aghanim, N. and others",
    collaboration = "Planck",
    title = "{Planck 2018 results. VI. Cosmological parameters}",
    eprint = "1807.06209",
    archivePrefix = "arXiv",
    primaryClass = "astro-ph.CO",
    doi = "10.1051/0004-6361/201833910",
    journal = "Astron. Astrophys.",
    volume = "641",
    pages = "A6",
    year = "2020",
    note = "[Erratum: Astron.Astrophys. 652, C4 (2021)]"
}

@article{DESI:2025zgx,
    author = "Abdul Karim, M. and others",
    collaboration = "DESI",
    title = "{DESI DR2 results. II. Measurements of baryon acoustic oscillations and cosmological constraints}",
    eprint = "2503.14738",
    archivePrefix = "arXiv",
    primaryClass = "astro-ph.CO",
    reportNumber = "FERMILAB-PUB-25-0169-PPD",
    doi = "10.1103/tr6y-kpc6",
    journal = "Phys. Rev. D",
    volume = "112",
    number = "8",
    pages = "083515",
    year = "2025"
}

@article{Coulton:2018ebd,
    author = {Coulton, William R. and Liu, Jia and Madhavacheril, Mathew S. and B{\"o}hm, Vanessa and Spergel, David N.},
    title = "{Constraining Neutrino Mass with the Tomographic Weak Lensing Bispectrum}",
    eprint = "1810.02374",
    archivePrefix = "arXiv",
    primaryClass = "astro-ph.CO",
    doi = "10.1088/1475-7516/2019/05/043",
    journal = "JCAP",
    volume = "05",
    pages = "043",
    year = "2019"
}

@article{Zhang:2019ipd,
    author = "Zhang, Jing-Fei and Wang, Bo and Zhang, Xin",
    title = "{Forecast for weighing neutrinos in cosmology with SKA}",
    eprint = "1907.00179",
    archivePrefix = "arXiv",
    primaryClass = "astro-ph.CO",
    doi = "10.1007/s11433-019-1516-y",
    journal = "Sci. China Phys. Mech. Astron.",
    volume = "63",
    number = "8",
    pages = "280411",
    year = "2020"
}

@article{Tanseri:2022zfe,
    author = "Tanseri, Isabelle and Hagstotz, Steffen and Vagnozzi, Sunny and Giusarma, Elena and Freese, Katherine",
    title = "{Updated neutrino mass constraints from galaxy clustering and CMB lensing-galaxy cross-correlation measurements}",
    eprint = "2207.01913",
    archivePrefix = "arXiv",
    primaryClass = "astro-ph.CO",
    doi = "10.1016/j.jheap.2022.07.002",
    journal = "JHEAp",
    volume = "36",
    pages = "1--26",
    year = "2022"
}

@article{Zhang:2015uhk,
    author = "Zhang, Xin",
    title = "{Impacts of dark energy on weighing neutrinos after Planck 2015}",
    eprint = "1511.02651",
    archivePrefix = "arXiv",
    primaryClass = "astro-ph.CO",
    doi = "10.1103/PhysRevD.93.083011",
    journal = "Phys. Rev. D",
    volume = "93",
    number = "8",
    pages = "083011",
    year = "2016"
}

@article{Zhao:2016ecj,
    author = "Zhao, Ming-Ming and Li, Yun-He and Zhang, Jing-Fei and Zhang, Xin",
    title = "{Constraining neutrino mass and extra relativistic degrees of freedom in dynamical dark energy models using Planck 2015 data in combination with low-redshift cosmological probes: basic extensions to {\ensuremath{\Lambda}}CDM cosmology}",
    eprint = "1608.01219",
    archivePrefix = "arXiv",
    primaryClass = "astro-ph.CO",
    doi = "10.1093/mnras/stx978",
    journal = "Mon. Not. Roy. Astron. Soc.",
    volume = "469",
    number = "2",
    pages = "1713--1724",
    year = "2017"
}

@article{Zhang:2017rbg,
    author = "Zhang, Xin",
    title = "{Weighing neutrinos in dynamical dark energy models}",
    eprint = "1703.00651",
    archivePrefix = "arXiv",
    primaryClass = "astro-ph.CO",
    doi = "10.1007/s11433-017-9025-7",
    journal = "Sci. China Phys. Mech. Astron.",
    volume = "60",
    number = "6",
    pages = "060431",
    year = "2017"
}

@article{Li:2024qso,
    author = "Li, Tian-Nuo and Wu, Peng-Ju and Du, Guo-Hong and Jin, Shang-Jie and Li, Hai-Li and Zhang, Jing-Fei and Zhang, Xin",
    title = "{Constraints on Interacting Dark Energy Models from the DESI Baryon Acoustic Oscillation and DES Supernovae Data}",
    eprint = "2407.14934",
    archivePrefix = "arXiv",
    primaryClass = "astro-ph.CO",
    doi = "10.3847/1538-4357/ad87f0",
    journal = "Astrophys. J.",
    volume = "976",
    number = "1",
    pages = "1",
    year = "2024"
}

@article{Giare:2024gpk,
    author = "Giar{\`e}, William and Najafi, Mahdi and Pan, Supriya and Di Valentino, Eleonora and Firouzjaee, Javad T.",
    title = "{Robust preference for Dynamical Dark Energy in DESI BAO and SN measurements}",
    eprint = "2407.16689",
    archivePrefix = "arXiv",
    primaryClass = "astro-ph.CO",
    doi = "10.1088/1475-7516/2024/10/035",
    journal = "JCAP",
    volume = "10",
    pages = "035",
    year = "2024"
}

@article{Jiang:2024xnu,
    author = "Jiang, Jun-Qian and Pedrotti, Davide and da Costa, Simony Santos and Vagnozzi, Sunny",
    title = "{Nonparametric late-time expansion history reconstruction and implications for the Hubble tension in light of recent DESI and type Ia supernovae data}",
    eprint = "2408.02365",
    archivePrefix = "arXiv",
    primaryClass = "astro-ph.CO",
    doi = "10.1103/PhysRevD.110.123519",
    journal = "Phys. Rev. D",
    volume = "110",
    number = "12",
    pages = "123519",
    year = "2024"
}

@article{Yang:2024kdo,
    author = "Yang, Yuhang and Ren, Xin and Wang, Qingqing and Lu, Zhiyu and Zhang, Dongdong and Cai, Yi-Fu and Saridakis, Emmanuel N.",
    title = "{Quintom cosmology and modified gravity after DESI 2024}",
    eprint = "2404.19437",
    archivePrefix = "arXiv",
    primaryClass = "astro-ph.CO",
    doi = "10.1016/j.scib.2024.07.029",
    journal = "Sci. Bull.",
    volume = "69",
    pages = "2698--2704",
    year = "2024"
}

@article{Giare:2024smz,
    author = "Giar{\`e}, William and Sabogal, Miguel A. and Nunes, Rafael C. and Di Valentino, Eleonora",
    title = "{Interacting Dark Energy after DESI Baryon Acoustic Oscillation Measurements}",
    eprint = "2404.15232",
    archivePrefix = "arXiv",
    primaryClass = "astro-ph.CO",
    doi = "10.1103/PhysRevLett.133.251003",
    journal = "Phys. Rev. Lett.",
    volume = "133",
    number = "25",
    pages = "251003",
    year = "2024"
}

@article{Ye:2024ywg,
    author = "Ye, Gen and Martinelli, Matteo and Hu, Bin and Silvestri, Alessandra",
    title = "{Hints of Nonminimally Coupled Gravity in DESI 2024 Baryon Acoustic Oscillation Measurements}",
    eprint = "2407.15832",
    archivePrefix = "arXiv",
    primaryClass = "astro-ph.CO",
    doi = "10.1103/PhysRevLett.134.181002",
    journal = "Phys. Rev. Lett.",
    volume = "134",
    number = "18",
    pages = "181002",
    year = "2025"
}

@article{Liu:2024yib,
    author = "Liu, Tonghua and Wang, Shengjia and Wu, Hengyu and Cao, Shuo and Wang, Jieci",
    title = "{Newest Measurements of Cosmic Curvature with BOSS/eBOSS and DESI DR1 Baryon Acoustic Oscillation Observations}",
    eprint = "2411.14154",
    archivePrefix = "arXiv",
    primaryClass = "astro-ph.CO",
    doi = "10.3847/2041-8213/adb7de",
    journal = "Astrophys. J. Lett.",
    volume = "981",
    number = "2",
    pages = "L24",
    year = "2025"
}

@article{Wolf:2025jed,
    author = "Wolf, William J. and Garc{\'\i}a-Garc{\'\i}a, Carlos and Anton, Theodore and Ferreira, Pedro G.",
    title = "{Assessing Cosmological Evidence for Nonminimal Coupling}",
    eprint = "2504.07679",
    archivePrefix = "arXiv",
    primaryClass = "astro-ph.CO",
    doi = "10.1103/jysf-k72m",
    journal = "Phys. Rev. Lett.",
    volume = "135",
    number = "8",
    pages = "081001",
    year = "2025"
}

@article{Pang:2025lvh,
    author = "Pang, Ye-Huang and Zhang, Xue and Huang, Qing-Guo",
    title = "{The impact of the Hubble tension on the evidence for dynamical dark energy}",
    eprint = "2503.21600",
    archivePrefix = "arXiv",
    primaryClass = "astro-ph.CO",
    doi = "10.1007/s11433-025-2713-8",
    journal = "Sci. China Phys. Mech. Astron.",
    volume = "68",
    number = "8",
    pages = "280410",
    year = "2025"
}

@article{Giare:2024oil,
    author = "Giar{\`e}, William",
    title = "{Dynamical dark energy beyond Planck? Constraints from multiple CMB probes, DESI BAO, and type-Ia supernovae}",
    eprint = "2409.17074",
    archivePrefix = "arXiv",
    primaryClass = "astro-ph.CO",
    doi = "10.1103/ss37-cxhn",
    journal = "Phys. Rev. D",
    volume = "112",
    number = "2",
    pages = "023508",
    year = "2025"
}

@article{Liu:2025mub,
    author = "Liu, Tonghua and Li, Xiaolei and Wang, Jieci",
    title = "{Dynamical Dark Energy in the Crosshairs: A Joint Analysis with DESI, Type Ia Supernovae, and TDCOSMO Constraints}",
    eprint = "2504.21373",
    archivePrefix = "arXiv",
    primaryClass = "astro-ph.CO",
    doi = "10.3847/1538-4357/aded8b",
    journal = "Astrophys. J.",
    volume = "988",
    number = "2",
    pages = "243",
    year = "2025"
}

@article{Cheng:2025lod,
    author = "Cheng, Hanyu and Di Valentino, Eleonora and Escamilla, Luis A. and Sen, Anjan A. and Visinelli, Luca",
    title = "{Pressure parametrization of dark energy: first and second-order constraints with latest cosmological data}",
    eprint = "2505.02932",
    archivePrefix = "arXiv",
    primaryClass = "astro-ph.CO",
    reportNumber = "CA21106; CA21136",
    doi = "10.1088/1475-7516/2025/09/031",
    journal = "JCAP",
    volume = "09",
    pages = "031",
    year = "2025"
}

@article{Ozulker:2025ehg,
    author = {{\"O}z{\"u}lker, Emre and Di Valentino, Eleonora and Giar{\`e}, William},
    title = "{Dark Energy Crosses the Line: Quantifying and Testing the Evidence for Phantom Crossing}",
    eprint = "2506.19053",
    archivePrefix = "arXiv",
    primaryClass = "astro-ph.CO",
    month = "6",
    year = "2025"
}

@article{Li:2025ops,
    author = "Li, Jun-Xian and Wang, Shuang",
    title = "{Reconstructing dark energy with model independent methods after DESI DR2}",
    eprint = "2506.22953",
    archivePrefix = "arXiv",
    primaryClass = "astro-ph.CO",
    doi = "10.1140/epjc/s10052-025-15065-1",
    journal = "Eur. Phys. J. C",
    volume = "85",
    number = "11",
    pages = "1308",
    year = "2025"
}

@article{Khoury:2025txd,
    author = "Khoury, Justin and Lin, Meng-Xiang and Trodden, Mark",
    title = "{Apparent w{\ensuremath{<}}-1 and a Lower S8 from Dark Axion and Dark Baryons Interactions}",
    eprint = "2503.16415",
    archivePrefix = "arXiv",
    primaryClass = "astro-ph.CO",
    doi = "10.1103/w4qb-plk8",
    journal = "Phys. Rev. Lett.",
    volume = "135",
    number = "18",
    pages = "181001",
    year = "2025"
}

@article{Li:2025ula,
    author = "Li, Yun-He and Zhang, Xin",
    title = "{Cosmic sign-reversal: non-parametric reconstruction of interacting dark energy with DESI DR2}",
    eprint = "2506.18477",
    archivePrefix = "arXiv",
    primaryClass = "astro-ph.CO",
    doi = "10.1088/1475-7516/2025/12/018",
    journal = "JCAP",
    volume = "12",
    pages = "018",
    year = "2025"
}

@article{Smith:2025icl,
    author = {Smith, Adam and {\"O}z{\"u}lker, Emre and Di Valentino, Eleonora and van de Bruck, Carsten},
    title = "{Dynamical Dark Energy Meets Varying Electron Mass: Implications for Phantom Crossing and the Hubble Constant}",
    eprint = "2510.21931",
    archivePrefix = "arXiv",
    primaryClass = "astro-ph.CO",
    month = "10",
    year = "2025"
}

@article{Li:2025dwz,
    author = "Li, Tian-Nuo and Wu, Peng-Ju and Du, Guo-Hong and Yao, Yan-Hong and Zhang, Jing-Fei and Zhang, Xin",
    title = "{Exploring non-cold dark matter in the scenario of dynamical dark energy with DESI DR2 data}",
    eprint = "2507.07798",
    archivePrefix = "arXiv",
    primaryClass = "astro-ph.CO",
    doi = "10.1016/j.dark.2025.102068",
    journal = "Phys. Dark Univ.",
    volume = "50",
    pages = "102068",
    year = "2025"
}

@article{Lemos:2025bgy,
    author = "Lemos, Thais",
    title = "{Estimating the baryon fraction in the IGM from well-localized FRBs and DESI data}",
    eprint = "2507.17693",
    archivePrefix = "arXiv",
    primaryClass = "astro-ph.CO",
    doi = "10.1088/1475-7516/2025/11/060",
    journal = "JCAP",
    volume = "11",
    pages = "060",
    year = "2025"
}

@article{You:2025uon,
    author = "You, Changyu and Wang, Dan and Yang, Tao",
    title = "{Dynamical dark energy implies a coupled dark sector: Insights from DESI DR2 via a data-driven approach}",
    eprint = "2504.00985",
    archivePrefix = "arXiv",
    primaryClass = "astro-ph.CO",
    doi = "10.1103/f6v7-n9fr",
    journal = "Phys. Rev. D",
    volume = "112",
    number = "4",
    pages = "043503",
    year = "2025"
}

@article{Ling:2025lmw,
    author = "Ling, Jia-Le and Du, Guo-Hong and Li, Tian-Nuo and Zhang, Jing-Fei and Wang, Shao-Jiang and Zhang, Xin",
    title = "{Model-independent cosmological inference after the DESI DR2 data with improved inverse distance ladder}",
    eprint = "2505.22369",
    archivePrefix = "arXiv",
    primaryClass = "astro-ph.CO",
    doi = "10.1103/p8nz-djjm",
    journal = "Phys. Rev. D",
    volume = "112",
    number = "8",
    pages = "083528",
    year = "2025"
}

@article{Liu:2025myr,
    author = "Liu, Tonghua and Li, Xiaolei and Xu, Tengpeng and Biesiada, Marek and Wang, Jieci",
    title = "{Torsion cosmology in the light of DESI, supernovae and CMB observational constraints}",
    eprint = "2507.04265",
    archivePrefix = "arXiv",
    primaryClass = "astro-ph.CO",
    doi = "10.1140/epjc/s10052-025-15090-0",
    journal = "Eur. Phys. J. C",
    volume = "85",
    number = "11",
    pages = "1351",
    year = "2025"
}

@article{Li:2025htp,
    author = "Li, Tian-Nuo and Du, Guo-Hong and Wu, Peng-Ju and Qi, Jing-Zhao and Zhang, Jing-Fei and Zhang, Xin",
    title = "{Testing the cosmic distance duality relation with baryon acoustic oscillations and supernovae data}",
    eprint = "2507.13811",
    archivePrefix = "arXiv",
    primaryClass = "astro-ph.CO",
    doi = "10.1140/epjc/s10052-025-15012-0",
    journal = "Eur. Phys. J. C",
    volume = "85",
    number = "11",
    pages = "1354",
    year = "2025"
}

@article{Li:2025muv,
    author = "Li, Tian-Nuo and Du, Guo-Hong and Li, Yun-He and Li, Yichao and Ling, Jia-Le and Zhang, Jing-Fei and Zhang, Xin",
    title = "{Updated constraints on interacting dark energy: A comprehensive analysis using multiple CMB probes, DESI DR2, and supernovae observations}",
    eprint = "2510.11363",
    archivePrefix = "arXiv",
    primaryClass = "astro-ph.CO",
    month = "10",
    year = "2025"
}

@article{Wu:2025wyk,
    author = "Wu, Peng-Ju",
    title = "{Comparison of dark energy models using late-universe observations}",
    eprint = "2504.09054",
    archivePrefix = "arXiv",
    primaryClass = "astro-ph.CO",
    doi = "10.1103/mcgb-ntwr",
    journal = "Phys. Rev. D",
    volume = "112",
    number = "4",
    pages = "043527",
    year = "2025"
}

@article{Wu:2024faw,
    author = "Wu, Peng-Ju and Zhang, Xin",
    title = "{Measuring cosmic curvature with non-CMB observations}",
    eprint = "2411.06356",
    archivePrefix = "arXiv",
    primaryClass = "astro-ph.CO",
    doi = "10.1103/sn3q-q589",
    journal = "Phys. Rev. D",
    volume = "112",
    number = "6",
    pages = "063514",
    year = "2025"
}

@article{Wang:2024dka,
    author = "Wang, Hao and Piao, Yun-Song",
    title = "{Dark energy in light of DESI DR1 and Hubble tension}",
    eprint = "2404.18579",
    archivePrefix = "arXiv",
    primaryClass = "astro-ph.CO",
    doi = "10.1016/j.physletb.2026.140180",
    journal = "Phys. Lett. B",
    volume = "873",
    pages = "140180",
    year = "2026"
}

@article{Li:2025vuh,
    author = "Li, Tian-Nuo and Du, Guo-Hong and Zhou, Sheng-Han and Li, Yun-He and Zhang, Jing-Fei and Zhang, Xin",
    title = "{Robust evidence for dynamical dark energy in light of DESI DR2 and joint ACT, SPT, and Planck data}",
    eprint = "2511.22512",
    archivePrefix = "arXiv",
    primaryClass = "astro-ph.CO",
    doi = "10.1016/j.dark.2026.102254",
    journal = "Phys. Dark Univ.",
    volume = "52",
    pages = "102254",
    year = "2026"
}

@article{Specogna:2025guo,
    author = "Specogna, Enrico and Adil, Shahnawaz A. and Ozulker, Emre and Di Valentino, Eleonora and Nunes, Rafael C. and Akarsu, Ozgur and Sen, Anjan A.",
    title = "{Updated constraints on omnipotent dark energy: A comprehensive analysis with CMB and BAO data}",
    eprint = "2504.17859",
    archivePrefix = "arXiv",
    primaryClass = "gr-qc",
    doi = "10.1103/b7ht-lx26",
    journal = "Phys. Rev. D",
    volume = "113",
    number = "10",
    pages = "103549",
    year = "2026"
}

@article{Fazzari:2025lzd,
    author = "Fazzari, Elisa and Giar{\`e}, William and Di Valentino, Eleonora",
    title = "{Cosmographic Footprints of Dynamical Dark Energy}",
    eprint = "2509.16196",
    archivePrefix = "arXiv",
    primaryClass = "astro-ph.CO",
    doi = "10.3847/2041-8213/ae2917",
    journal = "Astrophys. J. Lett.",
    volume = "996",
    number = "1",
    pages = "L5",
    year = "2026"
}

@article{Zhang:2025lam,
    author = "Zhang, Zhuoming and Xu, Tengpeng and Chen, Yun",
    title = "{Dynamical Dark Energy and the Unresolved Hubble Tension: Multi-model Constraints from DESI 2025 and Other Probes}",
    eprint = "2512.07281",
    archivePrefix = "arXiv",
    primaryClass = "astro-ph.CO",
    doi = "10.3847/1538-4357/ae4738",
    journal = "Astrophys. J.",
    volume = "999",
    number = "2",
    pages = "248",
    year = "2026"
}

@article{Song:2025bio,
    author = "Song, Ji-Yu and Du, Guo-Hong and Li, Tian-Nuo and Wang, Ling-Feng and Qi, Jing-Zhao and Zhang, Jing-Fei and Zhang, Xin",
    title = "{Gravitational wave standard sirens from GWTC-3 combined with DESI DR2 and DESY5: A late-universe probe of the Hubble constant and dark energy}",
    eprint = "2511.12017",
    archivePrefix = "arXiv",
    primaryClass = "astro-ph.CO",
    doi = "10.1007/s11433-025-2888-0",
    journal = "Sci. China Phys. Mech. Astron.",
    volume = "69",
    number = "4",
    pages = "240413",
    year = "2026"
}

@article{Lee:2025pzo,
    author = "Lee, Dong Ha and Yang, Weiqiang and Di Valentino, Eleonora and Pan, Supriya and van de Bruck, Carsten",
    title = "{Shape of dark energy: Constraining its evolution with a general parametrization}",
    eprint = "2507.11432",
    archivePrefix = "arXiv",
    primaryClass = "astro-ph.CO",
    doi = "10.1103/z7y2-yvhg",
    journal = "Phys. Rev. D",
    volume = "113",
    number = "6",
    pages = "063554",
    year = "2026"
}

@article{Santos:2025wiv,
    author = "Santos, Felipe Bruno Medeiros dos and Morais, Jonathan and Pan, Supriya and Yang, Weiqiang and Di Valentino, Eleonora",
    title = "{A new window on dynamical dark energy: combining DESI-DR2 BAO with future gravitational wave observations}",
    eprint = "2504.04646",
    archivePrefix = "arXiv",
    primaryClass = "astro-ph.CO",
    doi = "10.1088/1475-7516/2026/07/022",
    journal = "JCAP",
    volume = "07",
    pages = "022",
    year = "2026"
}

@article{Kessler:2025kju,
    author = "Kessler, Daniel A. and Escamilla, Luis A. and Pan, Supriya and Di Valentino, Eleonora",
    title = "{One-parameter dynamical dark energy: Hints for oscillations}",
    eprint = "2504.00776",
    archivePrefix = "arXiv",
    primaryClass = "astro-ph.CO",
    doi = "10.1103/1x6w-4tnd",
    journal = "Phys. Rev. D",
    volume = "114",
    number = "6",
    pages = "063501",
    year = "2026"
}

@article{Li:2025owk,
    author = "Li, Tian-Nuo and Du, Guo-Hong and Li, Yun-He and Wu, Peng-Ju and Jin, Shang-Jie and Zhang, Jing-Fei and Zhang, Xin",
    title = "{Probing the sign-changeable interaction between dark energy and dark matter with DESI baryon acoustic oscillations and DES supernovae data}",
    eprint = "2501.07361",
    archivePrefix = "arXiv",
    primaryClass = "astro-ph.CO",
    doi = "10.1007/s11433-025-2771-5",
    journal = "Sci. China Phys. Mech. Astron.",
    volume = "69",
    number = "1",
    pages = "210413",
    year = "2026"
}

@article{Li:2026hwq,
    author = "Li, Xiaolei and Liu, Tonghua and Li, Tian-Nuo and Du, Guo-Hong and Shafieloo, Arman and Biesiada, Marek",
    title = "{Metastability in Emergent Dark Energy: A New Framework Confronting Cosmological Observations}",
    eprint = "2606.05723",
    archivePrefix = "arXiv",
    primaryClass = "astro-ph.CO",
    doi = "10.3847/2041-8213/ae5a36",
    journal = "Astrophys. J. Lett.",
    volume = "1001",
    number = "1",
    pages = "L21",
    year = "2026"
}

@article{Li:2026xaz,
    author = "Li, Tian-Nuo and Giar{\`e}, William and Du, Guo-Hong and Li, Yun-He and Di Valentino, Eleonora and Zhang, Jing-Fei and Zhang, Xin",
    title = "{Robust Preference for Dark Sector Interactions}",
    eprint = "2601.07361",
    archivePrefix = "arXiv",
    primaryClass = "astro-ph.CO",
    month = "1",
    year = "2026"
}

@article{Turner:2026vzp,
    author = "Turner, Hannah C. and Dhawan, Suhail and Vagnozzi, Sunny and Townsend, Alice and Massey, Peter",
    title = "{Stress-testing the spatially flat Universe: nonparametric spatial curvature determination from DESI DR2}",
    eprint = "2609.22470",
    archivePrefix = "arXiv",
    primaryClass = "astro-ph.CO",
    month = "9",
    year = "2026"
}

@article{Wang:2026wrk,
    author = "Wang, Jia-Qi and Cai, Rong-Gen and Guo, Zong-Kuan and Li, Yun-He and Wang, Shao-Jiang and Zhang, Xin",
    title = "{Nonminimally coupled quintessence with sign-switching interaction}",
    eprint = "2604.02204",
    archivePrefix = "arXiv",
    primaryClass = "astro-ph.CO",
    doi = "10.1103/4sl7-m1qc",
    journal = "Phys. Rev. D",
    volume = "114",
    number = "4",
    pages = "043514",
    year = "2026"
}

@article{Wang:2025znm,
    author = "Wang, Jia-Qi and Cai, Rong-Gen and Guo, Zong-Kuan and Wang, Shao-Jiang",
    title = "{Resolving the Planck-DESI tension by nonminimally coupled quintessence}",
    eprint = "2508.01759",
    archivePrefix = "arXiv",
    primaryClass = "astro-ph.CO",
    doi = "10.1103/r6cx-8ghz",
    journal = "Phys. Rev. D",
    volume = "113",
    number = "8",
    pages = "083534",
    year = "2026"
}

@article{Yang:2025gaz,
    author = "Yang, Yuhang and Wang, Qingqing and Ren, Xin and Saridakis, Emmanuel N. and Cai, Yi-Fu",
    title = "{Probing the sound speed and clustering of dark energy}",
    eprint = "2511.22478",
    archivePrefix = "arXiv",
    primaryClass = "astro-ph.CO",
    doi = "10.1016/j.scib.2026.08.004",
    journal = "Sci. Bull.",
    volume = "71",
    pages = "4410--4415",
    year = "2026"
}

@article{Wang:2026vqw,
    author = "Wang, Jincheng and Yu, Hongwei and Wu, Puxun",
    title = "{Coupled quintessence with a potential from supergravity exhibits sign-changing interaction}",
    eprint = "2605.17754",
    archivePrefix = "arXiv",
    primaryClass = "astro-ph.CO",
    month = "5",
    year = "2026"
}

@article{Yang:2026qmz,
    author = "Yang, Weiqiang and Zhang, Sibo and Barbosa, Dinorah and Alvarez, Marco Antonio and Pan, Supriya and Graef, Leila",
    title = "{Do DESI-DR2 BAO data imply a coupling of dark matter and dark energy?}",
    eprint = "2609.05660",
    archivePrefix = "arXiv",
    primaryClass = "astro-ph.CO",
    doi = "10.1103/6kdf-vzq4",
    journal = "Phys. Rev. D",
    volume = "114",
    number = "6",
    pages = "063549",
    year = "2026"
}

@article{Pan:2025qwy,
    author = "Pan, Supriya and Paul, Sivasish and Saridakis, Emmanuel N. and Yang, Weiqiang",
    title = "{Interacting dark energy after DESI DR2: A challenge for the {\ensuremath{\Lambda}}CDM paradigm?}",
    eprint = "2504.00994",
    archivePrefix = "arXiv",
    primaryClass = "astro-ph.CO",
    doi = "10.1103/5y21-k39n",
    journal = "Phys. Rev. D",
    volume = "113",
    number = "2",
    pages = "023515",
    year = "2026"
}

@article{Yang:2026yaq,
    author = "Yang, Weiqiang and Di Valentino, Eleonora and Linder, Eric V. and Zhang, Sibo and Pan, Supriya",
    title = "{When one-parameter dark energy makes neutrinos physical again}",
    eprint = "2603.15422",
    archivePrefix = "arXiv",
    primaryClass = "astro-ph.CO",
    doi = "10.1103/41tc-5ffp",
    journal = "Phys. Rev. D",
    volume = "114",
    number = "6",
    pages = "063515",
    year = "2026"
}

@article{Du:2025xes,
    author = "Du, Guo-Hong and Li, Tian-Nuo and Wu, Peng-Ju and Zhang, Jing-Fei and Zhang, Xin",
    title = "{Cosmological preference for a positive neutrino mass at 2.7{\ensuremath{\sigma}}: A joint analysis of DESI DR2, DESY5, and DESY1 data}",
    eprint = "2507.16589",
    archivePrefix = "arXiv",
    primaryClass = "astro-ph.CO",
    doi = "10.1007/s11433-026-3051-4",
    journal = "Sci. China Phys. Mech. Astron.",
    volume = "69",
    number = "10",
    pages = "100413",
    year = "2026"
}

@article{Pedrotti:2026ncd,
    author = "Pedrotti, Davide and Giar{\`e}, William and Cheng, Hanyu and Di Valentino, Eleonora",
    title = "{When, Why, and How CMB Compression Fails}",
    eprint = "2610.08728",
    archivePrefix = "arXiv",
    primaryClass = "astro-ph.CO",
    month = "10",
    year = "2026"
}

@article{Yao:2025twv,
    author = "Yao, Yan-Hong and Shen, Yi-Hao and Li, Tian-Nuo and Du, Guo-Hong and Gong, Yungui",
    title = "{Examining a new form of non-standard dark matter using DESI DR2 data}",
    eprint = "2510.13436",
    archivePrefix = "arXiv",
    primaryClass = "astro-ph.CO",
    doi = "10.1016/j.dark.2026.102318",
    journal = "Phys. Dark Univ.",
    volume = "52",
    pages = "102318",
    year = "2026"
}

@article{Yang:2026vti,
    author = "Yang, Yupeng",
    title = "{Constraints on redshift-evolving Hubble constant models with early- and late-time diagnostics}",
    eprint = "2608.23589",
    archivePrefix = "arXiv",
    primaryClass = "physics.gen-ph",
    doi = "10.1140/epjp/s13360-026-08515-6",
    journal = "Eur. Phys. J. Plus",
    volume = "141",
    number = "11",
    pages = "1277",
    year = "2026"
}

@article{Zhou:2026dti,
    author = "Zhou, Sheng-Han and Li, Tian-Nuo and Du, Guo-Hong and Zhang, Yi-Min and Li, Zhao-Yu and Zhang, Jing-Fei and Zhang, Xin",
    title = "{Neutrino mass constraints in the Schwarzschild-de Sitter black-hole dark energy model with ACT DR6 and DESI DR2 data}",
    eprint = "2607.03183",
    archivePrefix = "arXiv",
    primaryClass = "astro-ph.CO",
    month = "7",
    year = "2026"
}

@article{Xu:2026sbw,
    author = {Xu, Tengpeng and Kumar, Suresh and Chen, Yun and Capistrano, Abra{\~a}o J. S. and Akarsu, {\"O}zg{\"u}r},
    title = "{Probing dynamical dark energy with late-time data: Evidence, tensions, and the limits of the w $_{0}$ waCDM framework}",
    eprint = "2602.11936",
    archivePrefix = "arXiv",
    primaryClass = "astro-ph.CO",
    doi = "10.1088/1475-7516/2026/08/004",
    journal = "JCAP",
    volume = "08",
    pages = "004",
    year = "2026"
}

@article{Zhang:2026axl,
    author = "Zhang, Ji-Guo and Li, Tian-Nuo and Du, Guo-Hong and Sun, Wan-Peng and Zhang, Jing-Fei and Zhang, Xin",
    title = "{Joint Constraints on the Hubble Constant and Cosmic Baryon Density from Localized Fast Radio Bursts and DESI DR2 Baryon Acoustic Oscillations}",
    eprint = "2610.15536",
    archivePrefix = "arXiv",
    primaryClass = "astro-ph.CO",
    month = "10",
    year = "2026"
}

@article{Jiang:2026ngo,
    author = "Jiang, Jun-Qian and Shafieloo, Arman",
    title = "{Is Dark Matter Really Matter?}",
    eprint = "2608.04763",
    archivePrefix = "arXiv",
    primaryClass = "astro-ph.CO",
    doi = "10.3847/2041-8213/aeaa2a",
    journal = "Astrophys. J. Lett.",
    volume = "1010",
    number = "1",
    pages = "L12",
    year = "2026"
}

@article{Ling:2026awm,
    author = "Ling, Jia-Le and Wang, Shao-Jiang",
    title = "{GPAge: a Global Parameterization based on cosmic Age for some unified early-time and late-time solutions to the Hubble tension}",
    eprint = "2610.30961",
    archivePrefix = "arXiv",
    primaryClass = "astro-ph.CO",
    month = "10",
    year = "2026"
}

@article{Zhou:2026ive,
    author = "Zhou, Sheng-Han and Li, Tian-Nuo and Du, Guo-Hong and Zhang, Yi-Min and Li, Yun-He and Zhang, Jing-Fei and Zhang, Xin",
    title = "{New constraints on holographic dark energy with high-redshift galaxy observations from JWST}",
    eprint = "2610.35951",
    archivePrefix = "arXiv",
    primaryClass = "physics.gen-ph",
    month = "10",
    year = "2026"
}

@article{Li:2026mbh,
    author = "Li, Tian-Nuo and Giar{\`e}, William and Du, Guo-Hong and Di Valentino, Eleonora and Anchordoqui, Luis A. and Wang, Shao-Jiang and Zhang, Xin and Cai, Rong-Gen",
    title = "{A Positive Turn: Coupled Quintessence and Cosmological Neutrino Masses}",
    eprint = "2611.14584",
    archivePrefix = "arXiv",
    primaryClass = "astro-ph.CO",
    month = "11",
    year = "2026"
}

@article{Li:2026jgg,
    author = "Li, Zhao-Yu and Li, Tian-Nuo and Du, Guo-Hong and Zhou, Sheng-Han and Li, Yun-He and Zhang, Jing-Fei and Zhang, Xin",
    title = "{Constraints on interacting dark energy with a constant equation of state in light of DESI DR2: The role of the interaction term}",
    eprint = "2610.30784",
    archivePrefix = "arXiv",
    primaryClass = "astro-ph.CO",
    month = "10",
    year = "2026"
}

@article{Sun:2026kig,
    author = "Sun, Tian-Yang and Li, Tian-Nuo and Du, Guo-Hong and Wang, He and Zhang, Jing-Fei and Zhang, Xin",
    title = "{Assessing the consistency of the CPL dark-energy model with cosmological data: Simulation-calibrated neural posterior disagreement}",
    eprint = "2610.31044",
    archivePrefix = "arXiv",
    primaryClass = "astro-ph.CO",
    month = "10",
    year = "2026"
}

@article{Du:2026qtq,
    author = "Du, Guo-Hong and Li, Tian-Nuo and Yin, Lu and Zhou, Sheng-Han and Wang, Hao and Zhang, Jing-Fei and Zhang, Xin",
    title = "{Resolving the Hubble Tension in the Early Dark Energy Framework with JWST and DESI Data}",
    eprint = "2606.19090",
    archivePrefix = "arXiv",
    primaryClass = "astro-ph.CO",
    doi = "10.3847/2041-8213/ae9f47",
    journal = "Astrophys. J. Lett.",
    volume = "1008",
    number = "2",
    pages = "L62",
    year = "2026"
}

@article{Du:2026cly,
    author = "Du, Guo-Hong and Li, Tian-Nuo and Liu, Tonghua and Zhang, Jing-Fei and Zhang, Xin",
    title = "{Evidence for deviation in gravitational light deflection from general relativity at cosmological scales with KiDS-Legacy and CMB lensing}",
    eprint = "2602.03110",
    archivePrefix = "arXiv",
    primaryClass = "astro-ph.CO",
    doi = "10.1007/s11433-026-2987-7",
    journal = "Sci. China Phys. Mech. Astron.",
    volume = "69",
    number = "8",
    pages = "280411",
    year = "2026"
}

@article{Du:2025csv,
    author = "Du, Guo-Hong and Li, Tian-Nuo and Ling, Jia-Le and Yao, Yan-Hong and Zhang, Jing-Fei and Zhang, Xin",
    title = "{Model-independent late-Universe measurements of H0 and {\ensuremath{\Omega}}K with the parametrization based on cosmic age-improved inverse distance ladder}",
    eprint = "2510.26355",
    archivePrefix = "arXiv",
    primaryClass = "astro-ph.CO",
    doi = "10.1103/rl22-1dyx",
    journal = "Phys. Rev. D",
    volume = "114",
    number = "4",
    pages = "L041304",
    year = "2026"
}

@article{Yang:2026qon,
    author = "Yang, Weiqiang and Zhang, Sibo and Pan, Supriya and Anchordoqui, Luis A. and Di Valentino, Eleonora",
    title = "{Exploring the Limits: Neutrino Mass and Interacting Dark Sector in the DESI Era}",
    eprint = "2610.33729",
    archivePrefix = "arXiv",
    primaryClass = "hep-ph",
    month = "10",
    year = "2026"
}

@article{Pulido-Hernandez:2026hcs,
    author = "Pulido-Hern{\'a}ndez, Hayyim and Cervantes-Cota, Jorge L.",
    title = "{Negative masses and spatial curvature: Effects on the neutrino mass tension in {\ensuremath{\Lambda}}CDM and extended cosmologies}",
    eprint = "2603.13208",
    archivePrefix = "arXiv",
    primaryClass = "astro-ph.CO",
    doi = "10.1103/c7j9-ykx6",
    journal = "Phys. Rev. D",
    volume = "114",
    number = "4",
    pages = "043543",
    year = "2026"
}

@article{Wang:2026gvg,
    author = "Wang, Jincheng and Yu, Hongwei and Wu, Puxun",
    title = "{Phantom-Divide Crossing in Exponentially Coupled Quintessence and the Role of Neutrino-Mass Freedom}",
    eprint = "2606.22023",
    archivePrefix = "arXiv",
    primaryClass = "astro-ph.CO",
    month = "6",
    year = "2026"
}

@article{Kibris:2026cqq,
    author = {K{\i}br{\i}s, Cihad and Elbers, Willem and Akarsu, {\"O}zg{\"u}r and Di Valentino, Eleonora},
    title = "{Negative neutrino mass or negative dark energy?}",
    eprint = "2605.21456",
    archivePrefix = "arXiv",
    primaryClass = "astro-ph.CO",
    month = "5",
    year = "2026"
}

@article{Zheng:2026tpt,
    author = "Zheng, Jie and Qiang, Da-Chun and You, Zhi-Qiang and Kumar, Darshan",
    title = "{Revisiting the equation of state of dark energy from DESI BAO with SNe Ia and CMB}",
    eprint = "2608.04353",
    archivePrefix = "arXiv",
    primaryClass = "astro-ph.CO",
    doi = "10.1140/epjc/s10052-026-16187-w",
    journal = "Eur. Phys. J. C",
    volume = "86",
    number = "8",
    pages = "924",
    year = "2026"
}

@article{CosmoVerseNetwork:2025alb,
    author = "Di Valentino, Eleonora and others",
    collaboration = "CosmoVerse Network",
    title = "{The CosmoVerse White Paper: Addressing observational tensions in cosmology with systematics and fundamental physics}",
    eprint = "2504.01669",
    archivePrefix = "arXiv",
    primaryClass = "astro-ph.CO",
    doi = "10.1016/j.dark.2025.101965",
    journal = "Phys. Dark Univ.",
    volume = "49",
    pages = "101965",
    year = "2025"
}

@article{Giare:2025pzu,
    author = "Giar{\`e}, William and Mahassen, Tariq and Di Valentino, Eleonora and Pan, Supriya",
    title = "{An overview of what current data can (and cannot yet) say about evolving dark energy}",
    eprint = "2502.10264",
    archivePrefix = "arXiv",
    primaryClass = "astro-ph.CO",
    doi = "10.1016/j.dark.2025.101906",
    journal = "Phys. Dark Univ.",
    volume = "48",
    pages = "101906",
    year = "2025"
}

@article{Zhang:2026zbs,
    author = "Zhang, Xin",
    title = "{Diagnosing dark energy in the post-DESI era: From effective dynamics to multi-probe consistency tests}",
    doi = "10.1007/s11433-026-3116-4",
    journal = "Sci. China Phys. Mech. Astron.",
    volume = "69",
    number = "12",
    pages = "120401",
    year = "2026"
}

@article{Cai:2025mas,
    author = "Cai, Yifu and Ren, Xin and Qiu, Taotao and Li, Mingzhe and Zhang, Xinmin",
    title = "{The quintom theory of dark energy after DESI DR2}",
    eprint = "2505.24732",
    archivePrefix = "arXiv",
    primaryClass = "astro-ph.CO",
    doi = "10.1093/nsr/nwag115",
    journal = "Natl. Sci. Rev.",
    volume = "13",
    number = "10",
    pages = "nwag115",
    year = "2026"
}

@article{Anchordoqui:2026hys,
    author = "Anchordoqui, Luis Alfredo and Lust, Dieter",
    title = "{Breaking Free from the Swampland of Impossible Universes Through the DESI Portal}",
    eprint = "2605.10476",
    archivePrefix = "arXiv",
    primaryClass = "astro-ph.CO",
    reportNumber = "MPP-2026-84",
    doi = "10.3390/universe12070216",
    journal = "Universe",
    volume = "12",
    number = "7",
    pages = "216",
    year = "2026"
}

@article{Li:2026asg,
    author = "Li, Tian-Nuo and Du, Guo-Hong and Wang, Hao and Li, Yun-He and Zhang, Jing-Fei and Zhang, Xin",
    title = "{Dark Energy in the DESI Era: A Brief Review of Evidence, Beyond-{\ensuremath{\Lambda}}CDM Interpretations, and Tensions}",
    eprint = "2606.21826",
    archivePrefix = "arXiv",
    primaryClass = "astro-ph.CO",
    doi = "10.1088/1674-4527/ae8429",
    journal = "Res. Astron. Astrophys.",
    volume = "26",
    number = "8",
    pages = "084002",
    year = "2026"
}

@article{Giare:2026oti,
    author = "Giar{\`e}, William and Lee, Dong Ha and Di Valentino, Eleonora",
    title = "{Intertwined constraints in extended cosmologies: Dark energy, curvature, neutrinos, and inflation}",
    eprint = "2607.01226",
    archivePrefix = "arXiv",
    primaryClass = "astro-ph.CO",
    doi = "10.1016/j.newar.2026.101759",
    journal = "New Astron. Rev.",
    volume = "103",
    pages = "101759",
    year = "2026"
}

@article{Chevallier:2000qy,
    author = "Chevallier, Michel and Polarski, David",
    title = "{Accelerating universes with scaling dark matter}",
    eprint = "gr-qc/0009008",
    archivePrefix = "arXiv",
    doi = "10.1142/S0218271801000822",
    journal = "Int. J. Mod. Phys. D",
    volume = "10",
    pages = "213--224",
    year = "2001"
}

@article{Linder:2002et,
    author = "Linder, Eric V.",
    title = "{Exploring the expansion history of the universe}",
    eprint = "astro-ph/0208512",
    archivePrefix = "arXiv",
    doi = "10.1103/PhysRevLett.90.091301",
    journal = "Phys. Rev. Lett.",
    volume = "90",
    pages = "091301",
    year = "2003"
}

@article{Brout:2022vxf,
    author = "Brout, Dillon and others",
    title = "{The Pantheon+ Analysis: Cosmological Constraints}",
    eprint = "2202.04077",
    archivePrefix = "arXiv",
    primaryClass = "astro-ph.CO",
    doi = "10.3847/1538-4357/ac8e04",
    journal = "Astrophys. J.",
    volume = "938",
    number = "2",
    pages = "110",
    year = "2022"
}

@article{DES:2024tys,
    author = "Abbott, T. M. C. and others",
    collaboration = "DES",
    title = "{The Dark Energy Survey: Cosmology Results with ~1500 New High-redshift Type Ia Supernovae Using the Full 5 yr Data Set}",
    journal = "Astrophys. J. Lett.",
    volume = "973",
    number = "1",
    pages = "L14",
    year = "2024",
    doi = "10.3847/2041-8213/ad6f9f"
}

@article{Jiang:2024viw,
    author = "Jiang, Jun-Qian and Giar{\`e}, William and Gariazzo, Stefano and Dainotti, Maria Giovanna and Di Valentino, Eleonora and Mena, Olga and Pedrotti, Davide and da Costa, Simony Santos and Vagnozzi, Sunny",
    title = "{Neutrino cosmology after DESI: tightest mass upper limits, preference for the normal ordering, and tension with terrestrial observations}",
    eprint = "2407.18047",
    archivePrefix = "arXiv",
    primaryClass = "astro-ph.CO",
    doi = "10.1088/1475-7516/2025/01/153",
    journal = "JCAP",
    volume = "01",
    pages = "153",
    year = "2025"
}

@article{Du:2024pai,
    author = "Du, Guo-Hong and Wu, Peng-Ju and Li, Tian-Nuo and Zhang, Xin",
    title = "{Impacts of dark energy on weighing neutrinos after DESI BAO}",
    eprint = "2407.15640",
    archivePrefix = "arXiv",
    primaryClass = "astro-ph.CO",
    doi = "10.1140/epjc/s10052-025-14094-0",
    journal = "Eur. Phys. J. C",
    volume = "85",
    number = "4",
    pages = "392",
    year = "2025"
}

\end{document}